\documentclass[sn-nature]{sn-jnl}
\usepackage{graphicx}%
\usepackage{multirow}%
\usepackage{amsmath,amssymb,amsfonts}%
\usepackage{amsthm}%
\usepackage{mathrsfs}%
\usepackage[title]{appendix}%
\usepackage{xcolor}%
\usepackage{textcomp}%
\usepackage{manyfoot}%
\usepackage{booktabs}%
\usepackage{algorithm}%
\usepackage{algorithmicx}%
\usepackage{algpseudocode}%
\usepackage{listings}%
\usepackage{chemformula}%
\usepackage{ulem}%
\usepackage{color}%

\usepackage{url}

\usepackage{hyperref}
\hypersetup{breaklinks=true}

\newcommand{\editor}[2]{%
  \expandafter\newcommand\csname #1note\endcsname[1]{%
    \textcolor{#2}{(\textbf{#1:} \textit{##1})}}%
  \expandafter\newcommand\csname #1\endcsname[1]{%
    \textcolor{#2}{##1}}%
  \expandafter\newcommand\csname #1cancel\endcsname[1]{%
    \textcolor{#2}{\sout{##1}}}%
  \expandafter\newcommand\csname #1change\endcsname[2]{%
    \textcolor{#2}{\sout{##1} ##2}}%
  \newenvironment{#1text}{\color{#2}}{\color{black}}
}
\editor{CE}{black}
\editor{HCN}{black}
\editor{FSH}{black}
\editor{KH}{black}
\editor{CK}{black}

\usepackage{xr}
\makeatletter
\newcommand*{\addFileDependency}[1]{
\typeout{(#1)}
\@addtofilelist{#1}
\IfFileExists{#1}{}{\typeout{No file #1.}}
}\makeatother

\DeclareUnicodeCharacter{03C0}{$\pi$}
\DeclareUnicodeCharacter{03C3}{$\sigma$}

\begin{document}

\title{Advances in momentum-resolved EELS of phonons, excitons and plasmons in 2D materials and their heterostructures}

\author[1]{\fnm{Cana} \sur{Elgvin}}
\author[1]{\fnm{Fredrik S.} \sur{Hage}}
\author[3]{\fnm{Khairi F.} \sur{Elyas}}
\author[3]{\fnm{Katja} \sur{Höflich}}
\author[1]{\fnm{Øystein} \sur{Prytz}}
\author[2,4]{\fnm{Christoph T.} \sur{Koch}}
\author[2,4]{\fnm{Hannah C.} \sur{Nerl}}

\affil[1]{\orgdiv{Department of Physics}, \orgname{University of Oslo}, \orgaddress{\street{Problemveien 11}, \city{Oslo}, \postcode{0313}, \country{Norway}}}

\affil[2]{\orgdiv{Department of Physics}, \orgname{Humboldt Universität zu Berlin}, \orgaddress{\street{Newtonstraße 15}, \city{Berlin}, \postcode{10587}, \country{Germany}}}

\affil[3]{\orgdiv{Ferdinand-Braun-Institut (FBH)},  \orgaddress{\street{Gustav-Kirchhoff-Str. 4}, \city{Berlin}, \postcode{12489},  \country{Germany}}}

\affil[4]{\orgdiv{Center for the Science of Materials Berlin}, \orgname{Humboldt Universität zu Berlin}, \orgaddress{\street{Newtonstraße 15}, \city{Berlin}, \postcode{10587}, \country{Germany}}}

\abstract{Functional nanomaterials, including 2D materials and their heterostructures are expected to impact fields ranging from catalysis, optoelectronics to nanophotonics. To realize their potential, novel experimental approaches need to be developed to characterize the combined materials and their components. Techniques using fast electrons, such as electron energy-loss spectroscopy (EELS), probe phenomena over an unrivaled energy range with high resolution. In addition, momentum-resolved EELS  simultaneously records energy and momentum transfer to the sample and thus generates two-dimensional data sets for each beam position. This allows excitations that occur at large momentum transfer to be resolved, including those outside of the light cone and beyond the first Brillouin zone, all whilst retaining nanometer sized spatial selectivity. \HCN{Such capabilities are particularly important} when probing phonons, plasmons, excitons and their coupling in 2D materials and their heterostructures.}


\maketitle
\newpage
\section{Introduction}\label{sec1-intro}
Since the discovery of graphene and the peculiar properties that are linked to its reduced dimensionality \cite{novoselov04_graphene}, interest in 2D materials for their (opto-)electronic properties has grown rapidly. Considering that in 2D materials (nearly) all atoms are exposed at the surfaces and the  concept from the more established field of semiconductors that "the interface is the device" \cite{kroemer_noble_lecture_2000}, exciting possibilities for discovering novel materials properties have emerged. Exploring the seemingly endless landscape of heterostructures produced by stacking different 2D materials on top of each other has opened up a new field of research focused on the creation of novel devices \cite{pham_2D_review_2022}.

Scanning transmission electron microscopy electron energy-loss spectroscopy (STEM-EELS) is an especially powerful technique for exploring the optoelectronic properties of 2D materials and their heterostructures, as it allows for the investigation of phenomena covering an unrivaled combination of energy (loss) range and resolution. The energy lost by the beam electrons covers a broad spectral range and thus phenomena with signatures from few $meV$ to a few thousand $eV$ can be investigated. Using a conventional STEM-EELS geometry, information on core transitions, bulk plasmons, excitons and phonons can be accessed by exciting them with an electron beam focused to a sub-Angstrom spot size.

\begin{figure}[h]
\centering
\includegraphics[width=0.8
\textwidth]{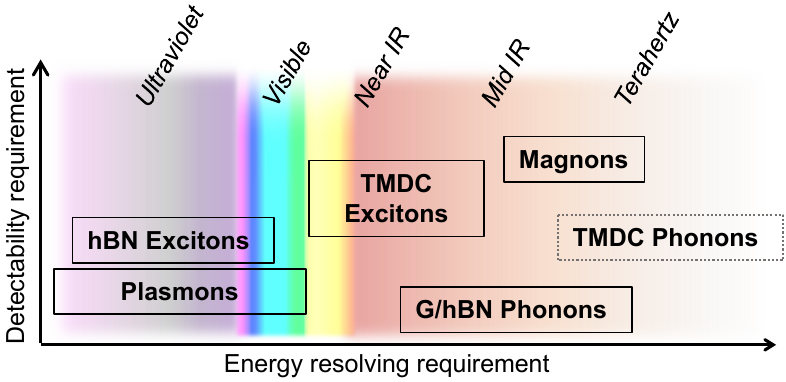}
\caption{Overview of phenomena that are accessible to probe in the electron microscope using momentum-resolved EELS, including excitons in hexagonal boron nitride (hBN) and transition metal dichalcogenides (TMDCs), plasmons, phonons in graphene (G) and hBN as well as in TMDCs, and magnons. The vertical axis, “Detectability requirement”, denotes the relative difficulty of detecting each excitation using EELS. Smaller cross sections and/or weaker signals lead to a higher “Detectability requirement”.
}\label{fig1-overviewGraph}
\end{figure}

While STEM-EELS necessitates integration over several Brillouin zones (BZs), momentum ($\hbar q$) - resolved EELS, from hereon referred to as q-EELS, gives direct access to the momentum transfer dependence of the energy-loss spectrum. Using q-EELS to investigate 2D materials, allows to spatially map the material at the nanoscale while providing simultaneous energy and momentum spectral information. Over the past years, this approach has been successfully used to study phenomena from plasmons to phonons as summarized in the schematic in Figure \ref{fig1-overviewGraph}.
In this perspective we briefly introduce the fundamentals of the technique and describe its resolution limits. In addition, detectability as a potential limitation will also be discussed.  We then summarize the current state of the art of q-EELS and survey the application of q-EELS for the study of a range of physical phenomena from plasmons to excitons to phonons in 2D materials as well as the specific challenges faced in heterostructures of 2D materials. We will then close our perspective with the main challenges that remain to be addressed and an outlook on what can be achieved in the foreseeable future. 
\section{Fundamentals of momentum-resolved EELS}\label{sec1-fundamentals}

\subsection{Physics of momentum-resolved EELS}\label{sec1-physics}

\begin{figure}[h]
\centering
\includegraphics[width=0.8\textwidth]{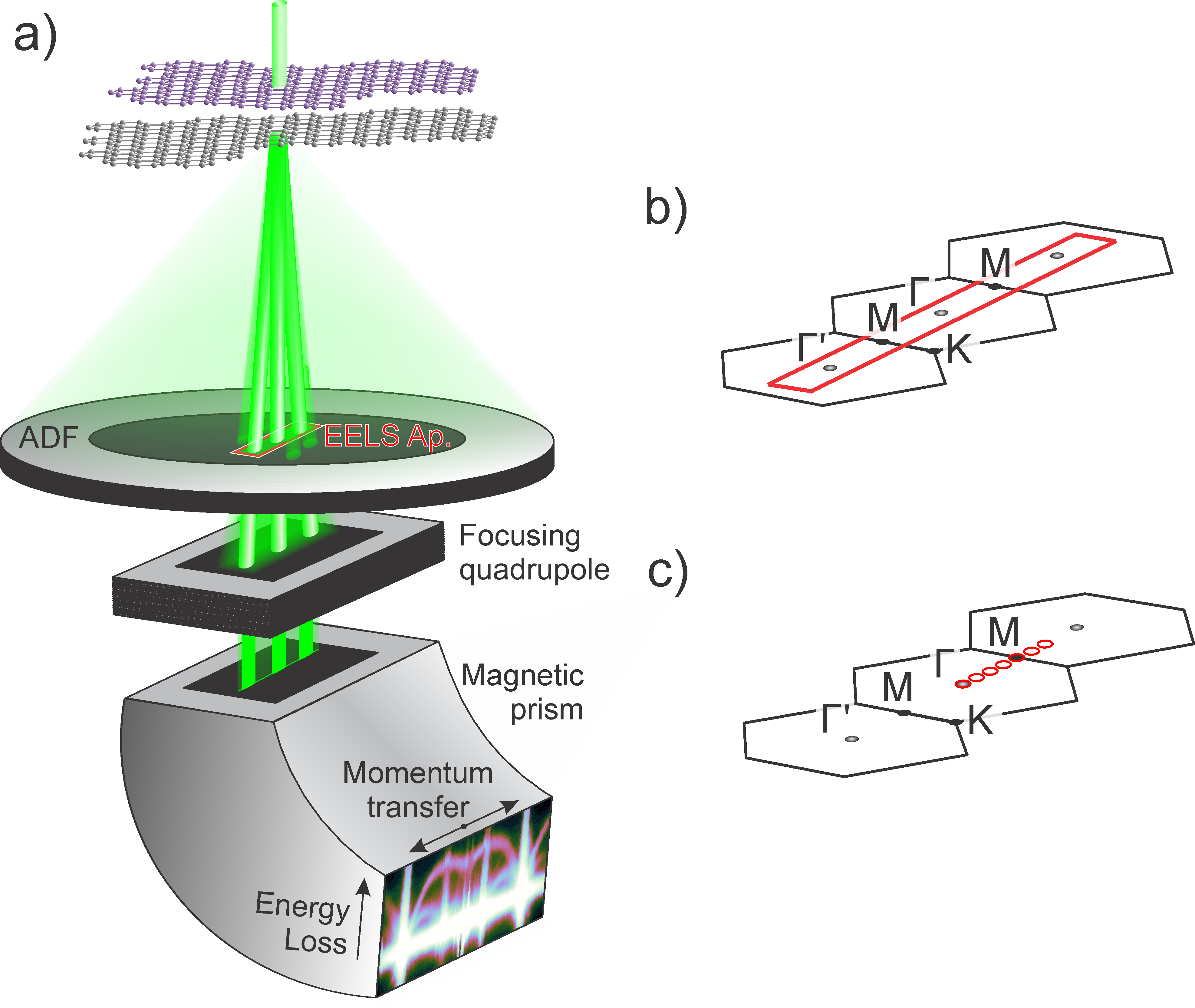}
\caption{Setup of the electron microscope for q-EELS. a) Schematic showing the slit aperture technique (denoted here with EELS aperture), including 
the annular dark field (ADF) detector, the focusing quadrupole, and the magnetic prism in the spectrometer. b) A slit aperture is aligned to high symmetry directions, e.g. along $\Gamma \rightarrow K$ and $\Gamma \rightarrow M$ to generate energy-momentum ($\omega q$) maps.  c) A circular aperture can  also be used, referred to as serial q-EELS, to successively acquire q-EEL spectra to map out the loss spectra at specific momentum transfers.}\label{fig2-setup}
\end{figure}

In STEM-EELS, the energy lost by the beam electron through interacting with the sample is measured as shown in Figure~\ref{fig2-setup}a and appears as peaks of certain intensities in the energy-loss spectrum. These peaks represent the probability of the creation of a certain excitation, where the beam electron is both the source of the excitation and its probe. The excitations being probed by the electron beam this way correspond to different phenomena within the material, such as electronic excitations, core shell transitions, plasmons, excitons, and phonons. 

Unlike freely propagating light, the electrons of the beam passing through or by the sample can be considered as moving point charges with an accompanying electromagnetic field that acts as an evanescent source of white (supercontinuum) light and can transfer nearly arbitrarily high momenta \cite{garciadeabajo2010optical,garcia2021optical}. Hence "everything" is pumped and probed at the same time within the limits of the energy and momentum provided by the electrons. \HCN{Here we use the terms pump and probe conceptually, since the fast electron acts as both the excitation source and the detector of the sample’s response via its evanescent field, rather than in the time-resolved sense of a pump--probe experiment as used in ultrafast spectroscopy.} This dual role of the electron underpins both the strength and the challenge of the technique: an observed feature at a given momentum in the loss spectrum arises from the coherent sum over all possible transitions from occupied to unoccupied states (and combinations thereof) that satisfy the corresponding momentum and energy transfer.

The total EEL spectrum $I(\omega, q)$ is specified by the double differential cross-section, which is most commonly expressed in the dielectric formulation. It is proportional to the loss function $\Gamma(\omega, q)$ as,
\begin{equation}
I(\omega,q) = \frac{d^2 \sigma(q,\omega)}{d\Omega \, dE} \;\propto\; \frac{1}{q^2}\,\Gamma(q,\omega),
\end{equation}
where the loss function is defined as
\[
\Gamma(q,\omega) = \mathrm{Im}\!\left[-\epsilon^{-1}(q,\omega)\right].
\]
The q-dependent loss function can be extracted from the experimentally obtained spectrum by processing steps, such as zero-loss peak subtraction \cite{reed2002zlp-deconvolution}, removal of multiple inelastic scattering by different Fourier-or iterative (e.g. Richardson-Lucy) deconvolution techniques \cite{Johnson1974eels-deconvolution,egerton_2011} and q-dependent normalization \cite{leon_unraveling_2024}. Momentum transfer $\mathbf{q}$ is related to the scattering angle as $q^2 \approx k_0^2(\theta^2 + \theta_E^2)$, where the characteristic angle $\theta_E$ can be neglected for small energy losses like those from plasmon, exciton and phonon excitations \cite{egerton_2011}.

We will refer to STEM-EELS as the on-axis setup typically used to record EELS spatially resolved across the sample, where a large convergence semi-angle $\alpha$ is employed to achieve high spatial resolution and maximum current on the detector. A diffraction pattern is projected in the spectrometer entrance plane, where a circular aperture is placed around the bright field disc center for maximum signal. The spectrometer then integrates the momentum information and disperses the electrons in their energy. Momentum-resolved EELS modifies this setup to access the momentum information available in the spectrometer entrance plane. There exist several methods to achieve this. 

Post-specimen deflectors can be used to select specific circular regions in the diffraction pattern by shifting the region of interest over the circular spectrometer entrance aperture. A two-dimensional momentum-resolved EELS map is then obtained by acquiring EEL spectra for successively shifted diffraction patterns (Fig. \ref{fig2-setup}c). 
Since almost every EEL spectrometer has circular entrance apertures, this technique, called serial q-EELS, is potentially available on most microscopes equipped with an EEL spectrometer. As the EELS intensity \CE{scales} with a $1/q^2$-dependence, this approach offers the flexibility to increase the acquisition time with increasing $q$. However, the exposure time should be recorded for each acquisition to allow for the recovery of the relative intensity variation with $q$.

In recent years, the use of a rectangular entrance slit to record 2D distributions of energy loss and momentum transfer, commonly referred to as $\omega q$ mapping and pioneered 50 years ago~\cite{Pettit1975}, has gained renewed importance (Fig. \ref{fig2-setup}a-b). This approach enables parallel acquisition of a continuous distribution of energy-losses across a range of momentum transfers, thereby reducing experimental time significantly compared to serial scans. \HCN{The diffraction pattern is oriented onto the slit by adjusting the projector lenses, with the long slit axis aligned parallel to the desired momentum-transfer direction and perpendicular to the energy-dispersive axis. The short dimension of the slit is integrated such that no momentum information is retained in the energy-dispersive direction.}

\subsection{Complementary Spectroscopic Techniques}\label{sec2-spectroscopy methods}

\HCN{Several established spectroscopies provide valuable and often higher-energy-resolution access to phonon and magnon dispersions, including inelastic neutron scattering (INS), non-resonant and resonant inelastic x-ray scattering (NRIXS and RIXS respectively, together often referred to simply as IXS), and high-resolution \textbf{reflectance} EELS (REELS). In the literature, the latter is often referred to as HREELS when emphasizing its millielectronvolt energy resolution. In this subsection we briefly compare their capabilities, sample requirements, and limitations with those of (\textbf{transmission}) q-EELS, to highlight the complementary information they provide.} \CE{Optical techniques such as Raman spectroscopy, photoluminescence and cathodoluminescence} \HCN{are, in turn, highly sensitive to zone-center ($q \approx 0$) optical phonons and excitons under well-defined selection rules, but cannot access finite-momentum excitations outside the light cone. This makes them complementary to q-EELS in terms of momentum coverage~\cite{ferrari2013raman,luo2020raman,mak+10prl,splendiani2010emerging,dang_cathodoluminescence_2023,tebyetekerwa2020PL}}

\CE{The principal advantage of EELS, and by extension q-EELS, is its superior spatial resolution which enables direct correlation of the dielectric (low loss) response with chemical composition and local geometrical information at the nanoscale.} \CE{By contrast, the x-ray and neutron-based spectroscopies (NR)IXS and INS also probe a response proportional to the dielectric function, but with different momentum-transfer dependencies and experimental constraints. In q-EELS, the momentum-dependent scattering cross section scales as $1/q^{2}$ while in (NR)IXS and INS it scales as $q^{2}$ ~\cite{foss+17prb,nicholls_theory_2019}. 
As a result, the latter suffer from a reduced signal at small momentum transfer (small $q$), but can more readily access higher BZs.} \CE{Their energy resolution is instrument dependent, but sub-meV energy resolution has been achieved in (NR)IXS and INS~\cite{IN12, cai_ultrahigh_2013, wang_non_resonant_2020}, which makes them dominant methods for measuring phonon and magnon dispersions.}

\HCN{INS offers the additional advantage of intrinsic spin sensitivity due to the neutron’s magnetic moment, while resonant inelastic x-ray scattering (RIXS), can access spin and orbital excitations via strong spin–orbit coupling at transition-metal absorption edges  ~\cite{furrer2009neutron,zhitomirsky2013colloquium,ament2011rixs}. Unlike NRIXS, however, RIXS spectra are not directly comparable to dielectric loss functions, which limits its complementarity to q-EELS.} \HCN{In the low $q$ regime, q-EELS provides superior performance by retaining sensitivity to excitations at small but finite momentum transfers, i.e. just outside the optical limit, where photon-based techniques cannot readily access.}

\HCN{Sample requirements also differ markedly. As neutrons interact only weakly with matter, their mean free path is on the order of centimeters. Consequently, INS requires relatively large volumes of high-quality single crystals (often on the order of several mm$^3$)  and long measurement times \CE{due to a low scattering cross-section} \cite{fultz_neutron_2013}. (NR)IXS can be applied to smaller crystals or polycrystalline samples without long range order, although the required sample sizes are still much larger than those needed for q-EELS. Both techniques depend on large-scale facilities, with limited availability and much longer measurement times compared to q-EELS}. \HCN{At the other end of the scale, REELS has been used since the 1960s to probe surface vibrational modes with meV or sub-meV resolution and intrinsic surface sensitivity~\cite{ibach_mills_1982,ibach1994electron}.}

\HCN{By controlling incidence and scattering angles, REELS accesses phonon dispersions in reflection geometry, and has been widely applied to systems such as graphene and adsorbates~\cite{eisinger_adsorption_1959, aizawa_phonon_1990}. (Transmission) q-EELS, in contrast, combines momentum resolution with nanoscale spatial mapping in thin electron-transparent specimens. Modern monochromated instruments now routinely achieve meV-scale energy resolution with high signal sensitivity, enabling advances such as the recent detection of magnons~\cite{kepaptsoglou2024magnon}.}

\HCN{Hence, the techniques are highly complementary in the samples and signals that can be probed: INS and (NR)IXS provide bulk-sensitive dispersions from large crystals, REELS offers surface-sensitive dispersions in reflection, and (transmission) q-EELS enables nanoscale mapping of excitations in thin samples.} \CE{An additional advantage of q-EELS is the} flexibility of STEM optics, which can be tuned to optimize for momentum, spatial and energy resolutions for a given experiment, and which allow to quickly switch between, or combine, imaging and diffraction modes. Moreover, q-EELS can be combined with spatial mapping, either by mapping targeted momentum transfers with an off-axis beam geometry \cite{hage_single-atom_2020}, or by acquiring a full four-dimensional hypercube that stores an $\omega q$ map at every beam position in the hyper-spectral dataset \cite{Wu2023FourdimensionalEE}.

\subsection{Resolution limits in momentum-resolved EELS}
\label{sec2}

\begin{figure}[h]
\centering
\includegraphics[width=1\textwidth]{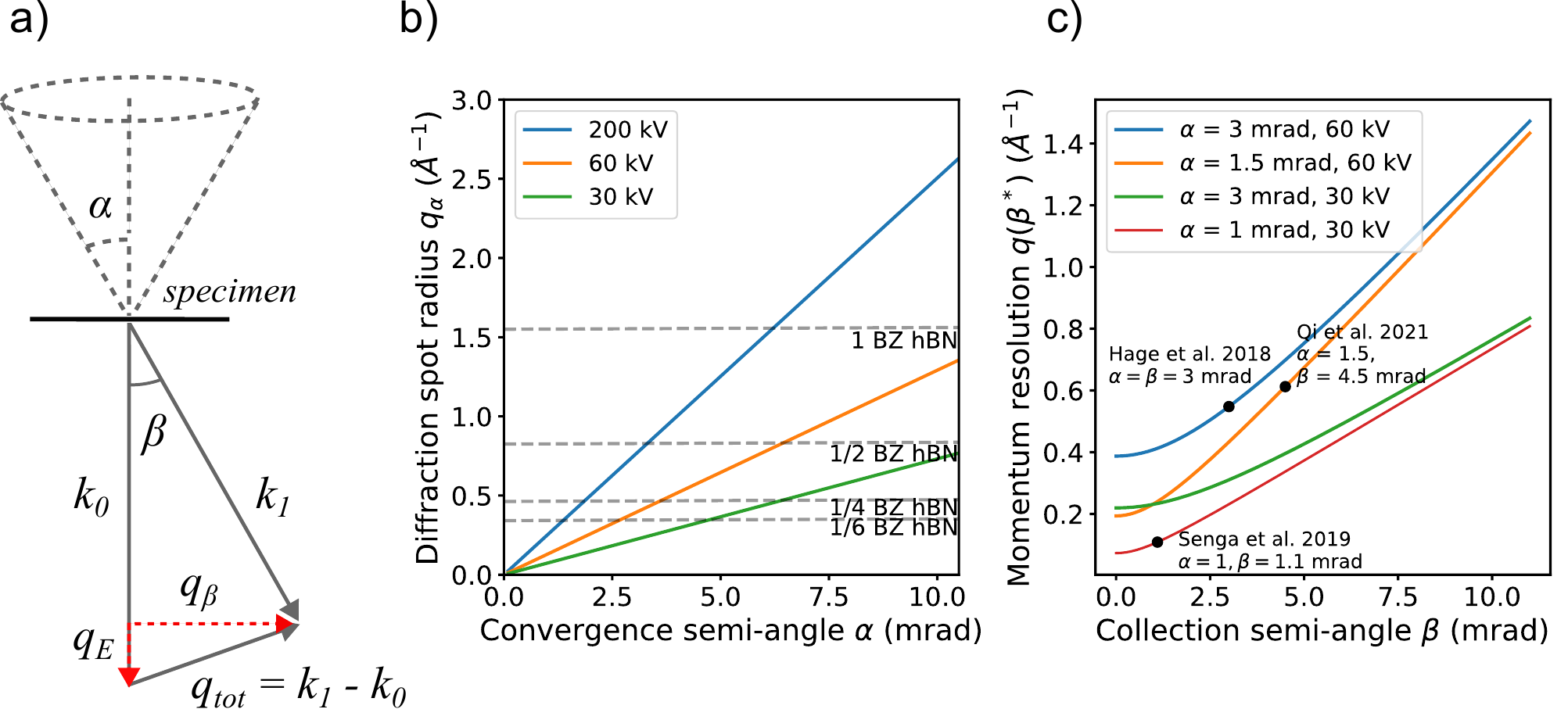}
\caption{\HCN{\textbf{(a)} Scattering geometry in momentum space. }\textbf{(b)}  Diffraction spot radius \CE{$\Delta q_\alpha$} as a function of convergence semi-angle $\alpha$, showing the intrinsic trade-off between momentum and spatial resolutions. Dashed lines indicate fractions of the Brillouin zone (BZ) of hBN along $\Gamma \rightarrow K$. \textbf{(c)} Momentum resolution specified as $q_{\beta^*} = (2\pi / \lambda) \sin(\beta^*)$ with $\beta^* = \sqrt{\beta^2 + \alpha^2}$. Examples of collection geometries from the literature are inserted from Hage \textit{et al.} \cite{hage_nanoscale_2018}, Senga \textit{et al.} \cite{senga_position_2019}, and Qi \textit{et al.} \cite{qi_four-dimensional_2021}. Note that \HCN{practical} resolution can be further limited by probe broadening and other instrumental factors.}\label{fig_geometry}
\end{figure}

In STEM, the diffraction limit constraints the simultaneously achievable spatial and momentum resolutions, i.e. the position-momentum uncertainty principle (see e.g., Ref. \cite{egerton2007limits}). A lower limit for the spatial resolution can be approximated by the diffraction-limited electron probe size in a scattering geometry as depicted in Figure \ref{fig_geometry}a. The solid lines in Figure \ref{fig_geometry}b illustrate this intrinsic trade-off between simultaneous real and momentum space resolution for $30 \,\mathrm{keV}$, $60\,\mathrm{keV}$ and $200\,\mathrm{keV}$ electrons, given by $\Delta q = (2 \pi / \lambda)\Delta \alpha$, where $\alpha$ is the convergence semi-angle \CE{and $\lambda$ the electron wavelength}. \CK{For values of $\alpha$ approaching the Scherzer angle, where probe aberrations start to play a significant role, the effective probe size is further limited by instrumental factors such as source size ( which typically increases with probe current) and residual aberrations \cite{williams_carter_2008, pennycook_nellist_2011}. While the diffraction-limited probe provides a lower limit for the spatial resolution, additional broadening arises from the delocalization of the inelastic interaction, especially at low energy losses and low momentum transfers (see discussion on inelastic delocalization below).}

\HCN{The total transferred momentum can be expressed in terms of  the convergence (incoming) and collection (accepted) angles as $q_{\text{tot}}^2 = q_\alpha^2 + q_\beta^2 + q^2_E$, where $\beta$ is the collection semi-angle, $q_E$ is the component of momentum transfer parallel to the incident beam direction, and  $q_{\alpha,\beta} = (2\pi / \lambda) \sin(\alpha,\beta)$. In the small-angle approximation, and under the common assumption that $q_E$ is negligible, we can define an effective collection aperture $\beta^* = \sqrt{\beta^2 + \alpha^2}$ }\CE{to specify the momentum resolution as $q_{\beta}^* = (2\pi / \lambda) \sin(\beta^*)$ for the scattering geometry shown in Fig. \ref{fig_geometry}a. \cite{egerton_2011}.}
Hence, limiting the convergence and collection angles \HCN{improves} the momentum resolution. \CE{When measuring q-EELS in the form of $\omega q$ mapping, a rectangular slit aperture is used instead of an annular aperture. The length of the slit is oriented in the $y$-direction, perpendicular to the energy-dispersive $x$-direction. Momentum selectivity is then given by the area $\Delta q_{x} \times \Delta q_{y}$ forming an integration window in reciprocal space, with increasing momentum transfers specified along $q_{y}$. In the case of parallel illumination ($\alpha \approx 0$), $\Delta q_{x}$ and $\Delta q_{y}$ determines the momentum resolution along the $x$ and $y$ axis, respectively.}

There are specific experimental designs that have been developed with the aim to maximize the momentum resolution. The "Midgley method" \cite{midgley1999} is based on changing the specimen height to produce a diffraction pattern in the intermediate image plane which integrates over the cone of illumination angles of the convergent illumination to then project a highly magnified image of this into the spectrometer entrance plane \cite{midgley1999, do_slow_light_2024}. In this way, the effective camera length can be chosen to be as large as several kilometers and the momentum resolution is not affected by the range of incident transverse wave vector components $q_\alpha$, since all those different angles of incidence are focused into a spot in the spectrometer entrance aperture. Another approach is to limit the collection angle to a minimum by inserting a small pin-hole aperture (at the cost of signal intensity) \cite{guandalini_excitonic_2023}. 

Depending on the aim of the experiment, a significant concern in q-EELS is often to maintain enough signal, as the intensity drops quadratically with finite \textit{q}. Hence, the requirement for signal intensity is higher than in STEM-EELS. This requirement must be balanced against the need for smaller spectrometer entrance apertures in q-EELS, since these select only a limited fraction of the potential energy-loss signal. Furthermore, to achieve the highest energy resolution, a monochromation by a factor of 15-20 is frequently required, leading to a reduced signal intensity.

\CE{So far we have considered physical and instrumental factors limiting the momentum and spatial resolution in q-EELS.}
We now turn to two important probe-sample interactions relevant to spatial resolution-- namely probe broadening and inelastic delocalization. 

Beam broadening refers to the electron beam being spread perpendicular to the forward direction by scattering within the sample. For $100\,\mathrm{keV}$ electrons scattered by a $50\,\mathrm{nm}$ thick amorphous carbon or gold film, elastic scattering broadens the probe by $\approx 2\,\mathrm{nm}$ and $\approx 20\,\mathrm{nm}$, respectively. Probe broadening in crystalline materials with the beam traveling along a low-index zone axis involves channeling (probe electrons are "guided along" to atomic columns) and is expected to lead to a lesser spread than for amorphous materials~\cite{egerton2007limits}. Probe propagation effects have been shown to give rise to non-intuitive contrast in STEM-EELS maps, for instance at the Si L$_{2,3}$ core loss edge \cite{wang2008contrast} and in vibrational loss maps of hexagonal boron nitride (hBN)~\cite{hage_contrast_2020}. \HCN{However, the impact of probe broadening can be mitigated by employing sufficiently thin specimens, as discussed in Ref.~\cite{egerton2007limits}.}

Inelastic delocalization may in practical terms be understood as the incoming electron beam having a significant probability of undergoing inelastic scattering within a region much larger than \HCN{the nominal probe size.} Classically, small impact parameters correspond to large scattering angles (and thus momentum transfers) and \textit{vice versa}. Within quantum mechanics, the Heisenberg uncertainty principle \HCN{states that position and momentum uncertainties are inversely related, i.e. $\Delta x \, \Delta p_x \gtrsim \hbar/2$. Thus, improved localization in the $x$-direction (smaller $\Delta x$)  necessarily leads to greater uncertainty in momentum $\Delta p_x$, and vice versa \cite{egerton_2011}.} 
\HCN{Therefore}, spectra acquired at higher $q$ in q-EELS experiments are associated with more localized (small  $\Delta x$) scattering events.

The degree of delocalization depends strongly on energy-loss: from the atomic scale for core losses to $\sim  0.1 \mu m$ for vibrational losses. Bohr's classical treatment~\cite{bohr1913constitution} of a fast free electron interacting with a bound electron gives the expression~\cite{muller1995delocalization, egerton2007limits, egerton_2011}:
\begin{equation} \label{eq:1}
\Delta E(b) =  \frac{2 e^4}{m v^2} \left( \frac{1}{b^2} \right)[K_{0}^2(b/b_{max}) + K_{1}^2(b/b_{max})]
\end{equation}

\noindent where $b$ is the impact parameter and $\Delta E (b)$ the corresponding energy loss. $K_{0}$ and $K_{1}$ are modified Bessel functions, and $b_{\max} = v/\omega$ is the adiabatic cut-off impact parameter, $v$ and $\omega$ being the velocity of the incoming electron and the resonance frequency of the bound electron, respectively \cite{jackson_classical_1999, muller1995delocalization, egerton2007limits, egerton_2011}. The time interval associated with the interaction is on the order of $\Delta t \sim \frac{b}{v}$ \cite{jackson_classical_1999}. 

For $b \ll b_{max}$ the interaction time is much shorter than the oscillation period of the bound electrons $\sim 1/\omega$. In this limit, the atomic electron only has time to behave much like a free electron during the scattering event. This means the interaction energy transfer can be described as Coulomb scattering between a stationary and a fast impinging electron, consistent with $\Delta E(b) \propto \frac{1}{b^2}$~\cite{jackson_classical_1999, muller1995delocalization, egerton2007limits, egerton_2011}. However, for $b \geq b_{max}$ the atomic electron "has time" to move resulting in a minimal energy transfer during the interaction or "dynamical screening", i.e., the interaction becomes adiabatic and $\Delta E(b) \propto \exp{\frac{-2b\omega}{v}}$~\cite{jackson_classical_1999, muller1995delocalization, egerton2007limits, egerton_2011}. While more sophisticated treatments have been developed and are necessary for a complete description of the physics involved, the classical picture above captures the main experimental trends (see e.g. \cite{egerton_2011} and references therein).

\begin{figure}[h]
\centering
\includegraphics[width=0.5\textwidth]{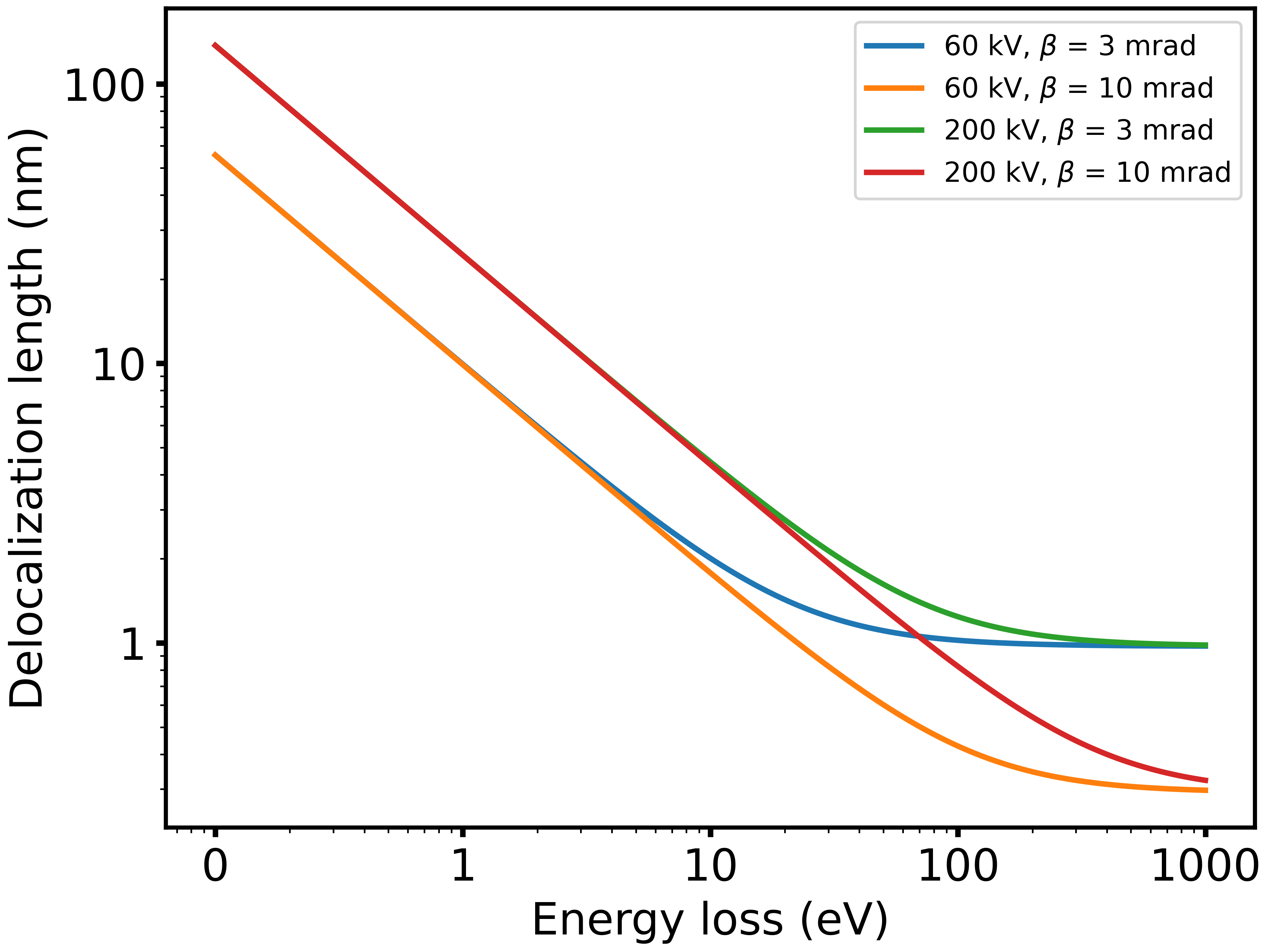}
\caption{\HCN{Delocalization length $d_{50}$ as a function of energy loss $\Delta E$ for two collection semi-angles, $\beta=3$ and $10$ mrad \cite{egerton2017scattering}. Smaller collection angles (q-EELS regime) cut off localized scattering at lower energy losses, while larger angles (common STEM-EELS regime) extend the range. This illustrates how spatial resolution depends strongly on geometry. In practice, }\CE{small, angle-limiting apertures in q-EELS can be used to traverse momentum space by isolating signal from delocalized small-angle scattering to more localized high-angle scattering.}}
\label{fig-delocalization}
\end{figure}

From Eq. \ref{eq:1} and the subsequent discussion, one would expect inelastic scattering to be localized in the (real space) specimen plane to a distance $L \approx b_{max}=v/\omega$. It can however be shown that a considerable fraction of inelastic scattering events are associated with impact parameters significantly smaller than $v/\omega$ \cite{egerton2007limits}. Even without angle-limiting spectrometer aperture, about half of the inelastic scattering is contained within \HCN{a disc of radius} $L_{50} \approx 0.6\lambda/ \langle \theta \rangle$, where $\langle \theta \rangle$ is the median scattering angle. For single electron transitions, this can be approximated further as $L_{50} \approx 0.5\lambda/\theta_{E}^{3/4}$, where $\theta_{E}$ is the characteristic scattering angle for energy-loss $E$. Finally, the influence of a finite collection angle $\beta $ (as in conventional STEM-EELS geometry) can be added in quadrature to give: 
\begin{equation}
(d_{50})^2 \approx (0.5\lambda/\theta_{E}^{3/4})^2 + (0.6\lambda /\beta)^2
\end{equation}

The 50\% localization length varies only weakly with accelerating voltage above $100$ kV, and is typically $1-10\,\mathrm{nm}$ for valence losses and about $100\,\mathrm{nm}$ for vibrational losses \cite{egerton2007limits, egerton_2011} (see Figure \ref{fig-delocalization}).

In q-EELS, detectors collect spectra over a small, selected range of available scattering angles using either bright field (BF, aperture overlaps with BF disc) or dark field (DF, aperture does not overlap with BF disc) geometries. 
Muller and Silcox \cite{muller1995delocalization} predicted and showed experimentally, that the DF plasmon loss signal in amorphous carbon is significantly more localized than the BF signal. More recently, a DF-EELS geometry was implemented by Dwyer \textit{et al.} \cite{dwyer_2016} to reach $\sim 1\,\mathrm{nm}$ spatial resolution in vibrational mapping of hBN. Extending this approach with a larger convergence angle, Hage \textit{et al.} acquired atomically resolved DF-EELS vibrational maps of hBN \cite{hage_phonon_2019} and graphene \cite{hage_single-atom_2020}.

In addition to the above-mentioned contributions of probe size, probe broadening in the specimen and detector momentum integration, the energy-momentum-dependent scattering cross-section must also be taken into consideration when evaluating the "effective localization" of a STEM-EEL spectral feature.     

\subsection{Contributions from surfaces and relativistic effects to the EEL spectrum}\label{sec-Cerenkov}

\HCN{Ideally, within the dielectric formalism the total EEL spectrum reduces to the single scattering distribution (SSD), i.e. the part of the spectrum arising from a single inelastic event, given by $\text{SSD}(q,\omega) \;\propto\; \frac{1}{q^{2}}\,\Gamma(q,\omega)$, where $\Gamma(q,\omega)$ is the loss function. In practice, however, additional contributions such as \v{C}erenkov radiation, surface plasmon resonances, and guided light modes (confined by total internal reflection rather than interface charge oscillations) must also be considered. }

\begin{figure}[h]
    \centering
    \includegraphics[width=1\textwidth]{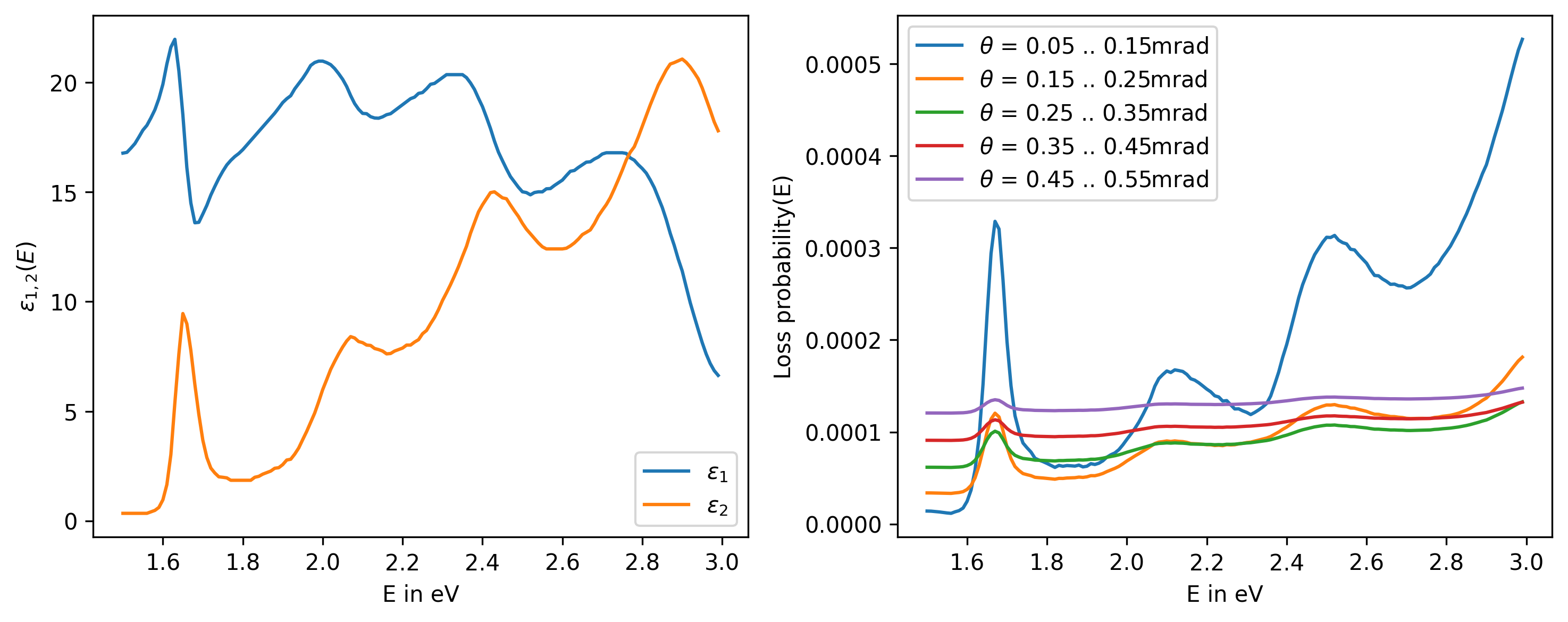}  \caption{a) Experimentally measured dielectric function of WSe$_2$ in the optical regime \cite{li_dielectric_function_TMDCs_2014}. b) Loss function including surface losses and relativistic effects computed from the dielectric function shown in (a) for different momentum transfers according to \cite{erni_relativistic_eels_2008}. Each spectrum is assumed to be collected for an annular aperture spanning the angular range specified in the legend.}
    \label{fig-relativistic}
\end{figure}

When the velocity of the electron beam exceeds the phase velocity of light in the dielectric material it passes through, it generates \v{C}erenkov radiation. Depending on the refractive index of the material, \v{C}erenkov radiation will, for example, be generated already at electron beam energies as low as 15 keV in the case of Germanium (Ge) ($n = 4.2$) \cite{gu_cerenkov_2007}. The dielectric constant of many TMDCs is even higher than that of Ge, suggesting that a contribution of \v{C}erenkov radiation would be present in practically any transmission EELS experiment performed on these 2D materials and their heterostructures. However, if the material is much thinner than the wavelength of the \v{C}erenkov radiation itself, it can often be neglected \cite{festenberg_cerenkov_1969,erni_relativistic_eels_2008,deabajo_cerenkov_2004}. For thicker specimens, however, \v{C}erenkov radiation produces spectral features in the optical regime of EELS,overlapping with the energy range of many opto-electronic phenomena. \v{C}erenkov radiation may, for example, interact strongly with excitons in these materials, leading to hybrid modes~\cite{Chahshouri_2022}.  If the angle of the \v{C}erenkov light cone is larger than the angle of total internal reflection relative to the direction of the electron beam, \v{C}erenkov radiation cannot escape the thin film and will thus propagate within it and contribute to guided light modes, enhancing the interaction with the phenomena under investigation~\cite{Chahshouri_2022}.  Since the excitation of \v{C}erenkov radiation limits the possible momentum transfer to the light line, it hardly deviates  from the forward direction. Figure  \ref{fig-relativistic}b shows the calculation of the loss function for $60\,\mathrm{kV}$ electron in a monolayer of WSe$_2$ using the Kröger formula \cite{kroeger_relativistic_eels_1968,erni_relativistic_eels_2008}. \HCN{In this model, the electrons are assumed to pass through a homogeneous, isotropic slab of thickness $t$ (here chosen as $0.5\,\mathrm{nm}$) characterized by a complex dielectric function $\epsilon(\omega) = \epsilon_{1}(\omega) + i \epsilon_{2}(\omega)$. This "local" approximation assumes that the dielectric response depends only on frequency and is independent of momentum transfer. In other words, the material is treated as homogeneous and isotropic in space, with dispersion in the optical sense (frequency dependence) but without spatial dispersion. While this approach is generally sufficient to capture bulk-like responses, it neglects nonlocal effects that become important in ultrathin 2D crystals where the full momentum and frequency-dependent dielectric function $\epsilon(\omega,q)$ must be considered.}

Recording the EELS signal off-axis or blocking the forward (BF) components of the electron beam in the entrance plane of the spectrometer allows for the contribution of \v{C}erenkov radiation generation to be blocked from the EEL spectrum \cite{gu_cerenkov_2007}. Using, for example, the Kröger formula (see Fig \ref{fig-relativistic}) to  compute contributions to the loss function due to the generation of \v{C}erenkov radiation or guided light modes, the dielectric function can then be refined in an iterative manner until the modelled EEL spectrum agrees with the experimental data, as has been demonstrated successfully for the case of a slab geometry that most TEM specimen come in \cite{eljarrat_relativistic_eels_2019}.

\HCN{In addition to \v{C}erenkov radiation, surface-related effects play a crucial role in low-loss EELS, particularly in ultrathin 2D crystals.  In such systems, the interface–electron interaction dominates: when an electron passes near the material boundary, it can excite confined surface plasmons, which strongly influence the low-loss response \cite{garciadeabajo2010optical}. Unlike bulk plasmons, which scale with the material thickness and volume electron density, surface plasmons are highly sensitive to boundaries, dielectric environment, and geometry. This sensitivity makes them particularly prominent in 2D materials and their heterostructures, where nearly all atoms lie at or near an interface.
The existence of surface plasmons in thin films was first predicted theoretically in 1957 \cite{ritchie1957plasma}. The first works to identify surface and bulk plasmons using EELS in aluminum were by Powell and Swan in 1959 \cite{powell1959origin}. Later, nanoscale EELS studies established surface-plasmon coupling in anisotropic hollow nanoparticles~\cite{kociak2001}, and the direct mapping of surface plasmons on single metallic nanoparticles by Nelayah \textit{et al.} demonstrated the spatially resolved character of these modes~\cite{nelayah2007}.}

\HCN{More recently (S)TEM EELS studies of topological insulator crystals revealed both, volume and surface plasmons with the latter observed at $\approx5.5 eV$ and $\approx10 eV $, which were distinct from bulk modes and visible in real-space spectral imaging~\cite{liou2013surface}. Furthermore, theoretical frameworks describe plasmon wakes (wake-field excitations trailing behind the fast electron) as well as directional plasmon excitation due to anisotropy and nonlocal effects, phenomena that are particularly relevant for anisotropic 2D materials such as phosphorene~\cite{miskovic2023plasmons}.
These studies have firmly established the conceptual distinction between surface and bulk plasmon modes, a distinction that becomes even more critical in 2D materials where the surface contribution dominates due to the ultrathin geometry.}

\HCN{The prominence of surface-plasmon signals in ultrathin films means that a correct interpretation of q-EELS data must disentangle these from \v{C}erenkov and bulk-loss contributions. Failing to include surface-plasmon effects can lead to misattribution of spectral features and incorrect conclusions about excitations in 2D systems. For example, in graphene and other metal-supported films, surface plasmons coexist and hybridize with excitonic excitations~\cite{Nerl2017ProbingTL, nerl2024mapping} (see Section\ref{sec-excitons}). Similarly, in ultrathin insulating layers such as hBN, surface phonon polaritons couple strongly to the electron beam,  dominating the low-loss spectra near $\Gamma$ and masking weaker bulk contributions (see Section\ref{sec-phonons}). 
\\
Importantly, the surface plasmon contribution modifies the momentum- and thickness-dependence of the scattering probability: while \v{C}erenkov losses become relevant at relativistic velocities, surface losses scale strongly with sample geometry and beam-sample coupling. Neglecting them would underestimate the low-loss background in ultrathin films. Thus, a full treatment of q-EELS in 2D materials must account for both \v{C}erenkov radiation and surface plasmon excitation in order to interpret spectral intensities correctly.}

\section{Applications of momentum-resolved EELS}
\subsection{Plasmons}\label{sec3-plasmons}
\HCN{(Bulk) plasmons are collective and coherent longitudinal oscillations of the valence electron density, sustained by long-range Coulomb interactions. In simple metals, they can be pictured as an in-phase motion of conduction electrons against the ionic background. In semiconductors, insulators, and 2D materials plasmons often involve interband transitions or hybridization with other excitations. The fundamental plasmon resonance, defined by the plasma frequency  $\omega_{p}$, corresponds to the natural collective density oscillation of the electron gas under Coulomb restoring forces.} Plasmons have important technological applications in fields such as optics, catalysis, solar energy conversion, and biosensing~\cite{maier2007plasmonics}. 

Probing the directional dependence of signals in momentum space at higher energies, such as plasmons, can be achieved using a range of electron microscopes. Due to their relatively high energies (several electronvolts in 2D materials and tens of electronvolts in bulk materials) and their substantial spectral width, there is no need to limit plasmon studies to instruments with the best resolving power. Indeed, investigations of the dispersion of plasmons represented the first q-EELS studies. \HCN{Watanabe already employed a q-slit to study the dispersion relation of bulk plasmons with q-EELS in 1956 \cite{watanabe_1956}. This came only a few years after Pines and Bohm had introduced the concept of the plasmon as a quantized quasiparticle~\cite{pines_bohm_1952}.  Within their random phase approximation (RPA) framework, they demonstrated that the collective response of a free electron gas arises from Coulomb interactions when perturbed by a fast charged particle. In this picture, a plasmon can be described as a coherent superposition of electron–hole transitions, involving predominantly intraband in metals and often involving interband contributions in semiconductors, or a mix of both depending on the material~\cite{pines1956collective,thyg17-2DM}.}
 
Plasmons are most commonly treated within the Lindhard's dielectric formulation, where the $q$-dependence of the bulk plasmon follows the parabolic Lindhard dispersion relation~\cite{lindhard_1954}.
\HCN{The condition for longitudinal plasmon modes follows directly from Maxwell’s equations: they occur when the real part of the dielectric function satisfies $\text{Re}\,\epsilon(q,\omega)=0$ with a positive slope. Experimentally, this condition is observed in EELS as a peak in the loss function $\mathrm{Im}[-1/\epsilon(q,\omega)]$. Beyond the fundamental plasmon at $\omega_p$, additional longitudinal modes exist at finite $q$ wherever the dielectric function vanishes, $\epsilon(q,\omega)=0$~\cite{maier2007plasmonics}. 
Dimensionality strongly modifies plasmon behavior. In three-dimensional (bulk) materials, efficient Coulomb screening ensures that plasmons dominate the low-loss spectrum. In contrast, the reduced dimensionality in 2D materials weakens long-range screening, which leads to a plasmon dispersion that scales as $\sqrt{q}$ in the long-wavelength limit~\cite{thyg17-2DM}. This behavior originates from the different scaling of the Coulomb interaction: $1/q^2$ in 3D versus $1/q$ in 2D. As a result, plasmon formation is strongly suppressed as $q \to 0$, but re-emerges at finite $q$ where Coulomb interactions dominate again over single electron–hole excitations.   This crossover highlights the fundamental difference between collective excitations in bulk and 2D systems. Crucially, the existence of longitudinal plasmon modes at finite $q$ makes momentum-resolved techniques such as q-EELS uniquely suited to probe them.  Optical spectroscopies in contrast, remain limited to probing near-zero momentum transfers within the light cone. }

The understanding of plasmons in 2D materials had been subject of much debate until it was elucidated with the help of q-EELS, through extensive studies of graphene and its related carbon systems in the 2010s. Graphene exhibits plasmonic $\pi$ and $\sigma$ peaks \HCN{similar to graphite, but their dispersions differ: graphite shows the expected parabolic dispersion of a 3D electron gas, while graphene reveals the $\sqrt{q}$ scaling characteristic of a 2D system.  Experiments and theory, however, have not always agreed on whether the $\pi$ plasmon dispersion is linear or follows $\sqrt{q}$. This reflects the difficulty of disentangling collective and single-particle contributions~\cite{lu+09prb, nelson_electronic_2014, liou_2015, novko2015, nazarov_2015}. Novko \textit{et al.} \cite{novko2015} resolved part of this controversy by demonstrating with DFT-RPA that the response is intrinsically mixed: the collective component follows $\sqrt{q}$ at intermediate $q$, while single-particle transitions dominate both at very small $q$ (quadratic dispersion linked to the band topology and weak screening) and at large $q$ (linear dispersion of the $\pi$ bands).} These theoretical results agreed with earlier q-EELS studies~\cite{wach+14PRB, kramberger2008linear, liou_2015}.  This shows that a non-generic momentum-dependence often reflects a mixed character,  which should be considered when interpreting peaks in q-EELS of 2D materials. Importantly, this case exemplifies how q-EELS has advanced our understanding of the nature of excitations. 

In other carbonic systems, early momentum-resolved measurements of bundled, single-walled carbon nanotubes (SWCT) showed a dispersive and non-dispersive plasmonic response \cite{pichler1998localized}. The dispersive response was interpreted to be confined to, and propagate along the tube axis \cite{pichler1998localized}. In contrast, non-dispersive (localized) interband peaks were assigned later to originating from van Hove singularities in the band structure. Thereafter, it was found that the $\pi$ (and $\pi+\sigma$) plasmons split into a non-dispersive (localized, $\pi_1$) and a linearly dispersive (delocalized, $\pi_2$) component \cite{kramberger2008linear}. The non-propagating $\pi_1$ mode was interpreted to be confined in the direction perpendicular to the tube axis, and the $\pi_2$ mode to propagate along it. The linear dispersion was only observed in the isolated SWCNTs, which suggests the presence of a mode confined to a one-dimensional wire. A non-propagating $\pi_1$ mode and a propagating $\pi_2$ mode were also measured for individual isolated SWCNTs by means of q-EELS \cite{hage2017nanotubes} while showing that structural SWCNT wall defects led to a disruption of the $\pi_2$ propagation. A confined $\pi$  mode was also observed in carbon nanocones \cite{hage2013nanocones}.

The momentum-dependence of the plasmon in 2D materials such as transition metal dichalcogenides (TMDCs)
has been discussed \cite{Nerl2017ProbingTL} using conventional STEM-EELS of \ch{MoS_2}, where the peaks in the plasmon region of the spectra were found to shift depending on thickness.
It was observed that when the beam traverses regions of \ch{MoS_2}, the plasmon peak was found to blue-shift with increasing thickness. This was explained by two effects: an increasingly larger momentum range being included in the measurements with increasing thickness such that a $q$-dependent contribution to the plasmon energy was detected and secondly that a higher electron density led to a higher plasmon energy. The former shows that a quantification of the thickness-dependence has to be interpreted with care due to the considerations mentioned above regarding sampling a varying $q$ range depending on specimen thickness. Using EELS in TEM, the $\sqrt{q}$-dependent dispersion of plasmons in \ch{MoS_2} was successfully shown by Köster \textit{et al.} \cite{koster2022evaluation} using a range of thicknesses. Later, the response of the plasmons of the TMDC \ch{WSe_2} was also shown to be isotropic along the high symmetry directions using q-EELS and observed to vary with thickness using DFT calculations \cite{nerl2024mapping}. 

The momentum-dependence of plasmons and excitons was also shown for \ch{PtSe_2} \cite{hong2022momentum}. In monolayer \ch{PtSe_2}, the effect of Coulomb screening on exciton and plasmon formation mentioned earlier, was clearly shown. It was found that the excitonic features dominate for $q \rightarrow 0$, while the plasmonic features grow dominant at high \textit{q}. The assignment of specific features to either excitonic or plasmonic origin is not trivial. Similarly to the discussion on graphene, Hong \textit{et al.} \cite{hong2022momentum} suggests the plasmons are of interband nature, transitioning from a single particle nature into plasmons. In the few-layer and bulk systems, the material transitions from being semiconducting to metallic, which makes the interpretation of the spectra even more challenging.

Some metallic TMDCs exhibit a negative plasmon dispersion, meaning that the energy carried by the wave (group velocity) travels in the opposite direction to the wave propagation (phase velocity) \cite{Shalaev_negative_2005}. This was shown experimentally in 2H-TaSe$_2$, 2H-TaS$_2$ and 2H-NbSe$_2$ \cite{schuster_negative_2009, van_wezel_effect_2011}, and calculated by Cudazzo \textit{et al.} \cite{cudazzo_2012}. The origin of this dispersion is attributed to non-local q-dependent screening from strong single particle contributions and the polarizable background of valence electrons \cite{cudazzo_2012, gjerding2017}.

\begin{figure}[htbp]
\centering
\includegraphics[width=1\textwidth]{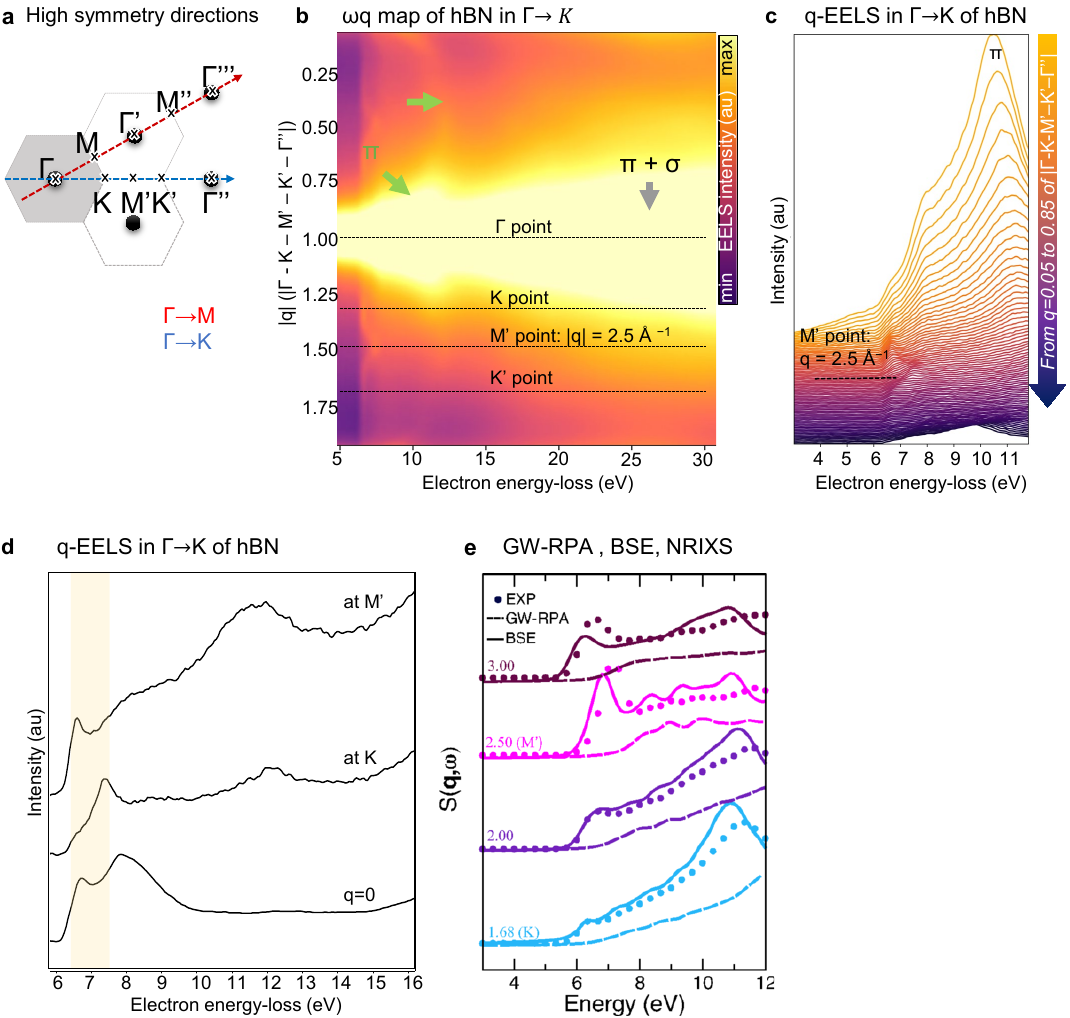}
\caption{Anisotropic dispersion of plasmons and excitons in hBN. \HCN{ \textbf{a}) Brillouin zone schematic showing the high-symmetry directions $\Gamma \rightarrow M$ (red) and $\Gamma \rightarrow K$ (blue).} \textbf{b}) Energy-momentum ($\omega q$) map in log scale along $\Gamma \rightarrow K$  from momentum-resolved EELS (q-EELS) and displayed in fractional units of momentum transfer where  $|\Gamma-\Gamma''|=1$. The $\pi$ plasmon peak was found to shift to higher energy with increasing \textit{$|q|$} (denoted with green arrow). The broad $\pi+\sigma$ plasmon peak is denoted by a grey arrow. The dispersion of both was shown to vary with thickness. \textbf{c}) Individual q-EEL spectra showing the spectral intensity redistribution of the excitonic intensities at the M' point at $|q| = 2.5 \text{\AA} ^{-1}$ along $\Gamma \rightarrow K$ which was found to be distinct from the spectral signatures along $\Gamma \rightarrow M$. 
\textbf{d}) Selected individual q-EEL spectra showing the spectral intensity redistribution of the excitonic intensities at the $M'$ point at $|q| = 2.5 \text{AA}^{-1}$ compared to $q=0$ and at the $K$ point along $\Gamma \rightarrow K$. e) The experimental results are compared with the calculated dynamic structure factor for \textit{q} along $\Gamma \rightarrow K$ obtained from the Bethe-Salpeter equation (BSE) as well as with non-resonant inelastic x-ray scattering (NRIXS) data. For reference, results from GW-RPA calculations \HCN{(using $G$, the single-particle Green’s function, and $W$, the screened Coulomb interaction, within the random-phase approximation)} are also shown.  (Panels \textbf{b}-\textbf{d} reproduced from reference \cite{nerl2024mapping} and panel \textbf{e}) reproduced from reference \cite{fugallo+15prb} containing experimental data from reference \cite{gala+11prb}).}\label{fig-hBN}
\end{figure}

While TMDCs are in-plane anisotropic, their plasmonic response seems to be highly isotropic \cite{Nerl2017ProbingTL, yue+17prb, nerl2024mapping}. In h-BN however, the plasmonic response has been shown to be anisotropic in the in-plane directions along $\Gamma \rightarrow M$ and $\Gamma \rightarrow K$ using x-ray techniques  \cite{gala+11prb, foss+17prb} and later confirmed using q-EELS \cite{nerl2024mapping}. \HCN{The high symmetry directions along $\Gamma \rightarrow M$ and $\Gamma \rightarrow K$ are  shown in Figure \ref{fig-hBN}a}. The $\omega q$ map in Figure \ref{fig-hBN}b,  and the individual spectra in Figure \ref{fig-hBN}c show the plasmonic response along $\Gamma \rightarrow K$ of hBN, which exhibits fine structure that was found to be absent along $\Gamma\rightarrow M$ \HCN{(only response along $\Gamma \rightarrow K$ shown here with full set available in \cite{nerl2024mapping})}.  The $\pi$ plasmon peak was found to shift to higher energy with increasing \textit{$|q|$} (denoted with green arrows in Figure \ref{fig-hBN}b). The dispersion of both major plasmonic peaks was shown to vary with thickness.  Furthermore, it was shown that the $\pi$ and $\pi+\sigma$ plasmon peaks exhibit variations in the relative spectral weight and dispersion behavior when comparing both directions. The variation has been ascribed to additional features with interband character due to the anisotropy in the band structure \cite{gala+11prb}. This will be discussed further in the Section\ref{sec-excitons} when examining excitons in hBN. 
\clearpage
\subsection{Excitons}\label{sec-excitons}
Excitons are quasiparticles that govern optical and electronic properties of semiconductors. \HCN{In extended three-dimensional systems (bulk semiconductors), an exciton is a bound state of an electron and a hole held together by their Coulomb attraction. In what follows, we refer to this as a 'bulk exciton'. In 2D semiconductors, by contrast, excitons are much more strongly bound and spatially confined, so their behavior differs strongly from the bulk case.}
Excitons in TMDCs specifically have caused particular interest as they exhibit strong exciton binding energies and therefore longer lifetimes. This is due to their two-dimensionality causing strong Coulomb attraction as a result of the reduced dielectric screening and additional geometrical confinement compared to their three-dimensional counterparts. In 2D materials, it has been predicted computationally that plasmon and exciton oscillator strengths deviate from those of bulk materials~\cite{thyg17-2DM}.   \HCN{For example, ab initio calculations and optical measurements indicate that direct-gap exciton binding energies in monolayer TMDCs typically range from $0.5$--$1\,\mathrm{eV}$~\cite{zhu2015exciton,wang2018colloquium}; two-photon photoluminescence excitation spectroscopy extracted a binding energy of $0.71 \pm 0.01\,\mathrm{eV}$ for monolayer WS$_2$~\cite{zhu2015exciton}. The enhanced binding also leads to extended radiative lifetimes: first-principles calculations predict intrinsic radiative lifetimes of $190$--$240\,\mathrm{fs}$ for excitons at zero centre-of-mass momentum and $0\,\mathrm{K}$, and the effective radiative lifetime increases roughly linearly with temperature from $1$--$10\,\mathrm{ps}$ at $4\,\mathrm{K}$ to $1$--$5\,\mathrm{ns}$ at room temperature~\cite{palummo2015exciton}. Hence, TMDCs exhibit rich excitonic properties even at room temperature~\cite{wilson1969transition,Wang2012ElectronicsAO,ugeda2014giant}, which enhance their potential for photonic and optoelectronic applications even further~\cite{o2009photonic,Mueller2018ExcitonPA,blais2020quantum}. In particular, longer radiative lifetimes are advantageous for devices such as light-emitting diodes, lasers, and exciton-based condensates, as they increase the probability of radiative recombination and enable stronger light–matter coupling.  The literature on excitons in TMDCs is vast and we refer the reader to one of the many excellent review articles on the subject~\cite{wang2018colloquium} while we will focus on conventional STEM-EELS and q-EELS studies from hereon.}

\HCN{Conventional STEM-EELS has been widely used to study excitons in TMDCs, with most studies focusing on the low-energy A and B excitons, which arise from spin-orbit splitting of the valence band in these materials~\cite{woo+23prb,tizei2015exciton,tizei2016electron,Nerl2017ProbingTL,bonnet2021nanoscale}. Such investigations have been carried out in both simple TMDCs and more complex systems, including twisted bilayers and heterostructures~\cite{gogoi2019layer,woo+23prb}, as well as in TMDCs combined with graphene or hBN~\cite{PhysRevMaterials.6.074005,woo2024engineering}.}
The A and B excitons in monolayer WSe$_2$ have been calculated to be  as large as 0.4~eV~\cite{Deilmann2019FinitemomentumEL}. Currently there is no consensus on the higher energy excitonic peak assignment, even for some of the most well-studied TMDCs and we will focus on one of the most well-studied TMDCs, WSe$_2$, to highlight the challenges.  The lowest energy and most prominent exciton peak in the EEL spectra of a TMDC is reliably assigned as the A exciton.  Most recently, Hong \textit{et al.} ~\cite{Hong2021}  however, showed that there is significant overlap between the A' Rydberg state and the B exciton in WSe$_2$. Hence, the second peak has overlapping contributions from the B exciton and the A' Rydberg state exciton. This shows that in general, great care has to be taken when interpreting the peaks in conventional STEM-EELS as there are overlapping peaks from excitons and their excited state excitons, the so-called Rydberg state excitons which cannot always be resolved using STEM-EELS. The Rydberg state excitons are thought to shift to lower energies with increasing thickness~\cite{PhysRevLett.113.076802}, which adds an additional layer of complexity to the interpretation the spectral information.
\begin{figure}[h]
\centering
\includegraphics[width=1\textwidth]{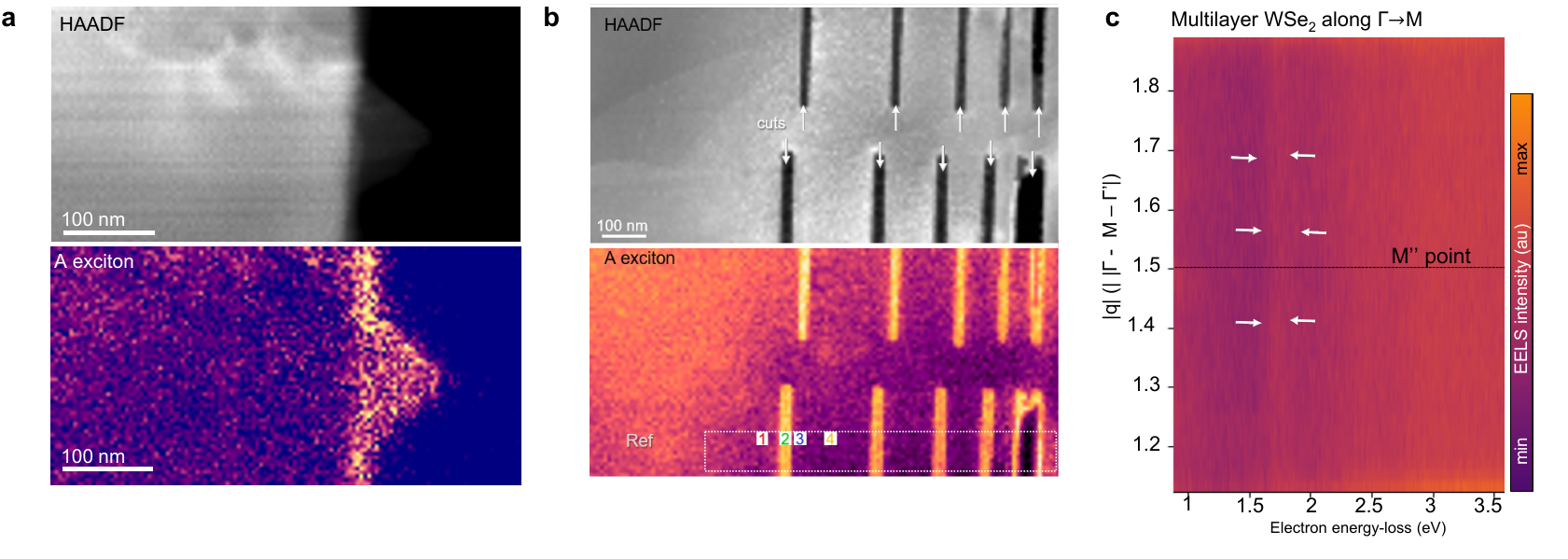}
\caption{At the onset of excitons polariton formation in transition metal dichalcogenide WSe$_2$: Conventional STEM-EELS and q-EELS show that excitonic signature peaks are likely to contain both, bulk exciton and exciton polariton contributions.  \textbf{a}) High angle annular dark field (HAADF) image showing a natural edge of WSe$_2$ (intensity scale: black= vacuum to white= thickest part of WSe$_2$). Below is the corresponding A exciton intensity map extracted by integrating the intensity of the A exciton peak in the STEM-EEL spectral map (scale in arbitrary units where purple=lowest intensity to yellow highest intensity). \HCN{Note that the A exciton is present across both monolayer and thicker regions; however, its relative intensity is maximized in the monolayer and diminishes with increasing layer number, while the energy remains nearly unchanged.} \textbf{b}) HAADF image showing WSe$_2$ with cuts from Helium Ion Microscopy (HIM) (intensity scale: black= vacuum to white= thickest part of the  WSe$_2$). Below is the corresponding A exciton intensity map from STEM-EELS (scale in arbitrary units where purple=lowest intensity to yellow highest intensity). Both, regions in a) and b) show that presence of edges leads to enhanced A excitonic intensities which cannot be explained by presence of bulk excitons alone. \textbf{c}) Energy-momentum map from q-EELS shows that at the onset, exciton polaritonic modes exhibit a near flat energy-momentum dispersion relation in $\approx12\,\mathrm{nm}$ thin films of WSe$_2$ along $\Gamma \rightarrow M$ across the entire Brillouin zone (arrows) (map displayed in fractional units of momentum transfer where  $|\Gamma-\Gamma'|=1$).  
Panel \textbf{a} shows previously unpublished data and panels \textbf{b}, \textbf{c} are reproduced from reference \cite{nerl2024flat}.}\label{fig-HIM-WeSe2}
\end{figure}
A major advantage of using conventional STEM-EELS to study excitonic intensities is the improved spatial resolution compared with optical techniques. Using this approach, excitonic intensities can be mapped out at the nanometer scale.\HCN{The intensity map of the A exciton (Figure \ref{fig-HIM-WeSe2}a) shows that the signal is strongly enhanced in the monolayer region protruding from the thicker flake. The A exciton is present in both monolayer and few-layer WSe$_2$, but in the latter its relative intensity decreases with increasing thickness due to enhanced dielectric screening, while its energy position remains nearly unchanged (within a few tens of meV). Hence, the map primarily reflects the thickness-dependent change in oscillator strength. In addition to enhanced excitonic intensity, edges are known to act as efficient launchers and confinement sites for polaritons, where broken symmetry and local field enhancement increase coupling strength and can modify dispersion relations.}
Nerl \textit{et al}. \cite{nerl2024flat} have used this approach to map out A excitonic intensities in nanopatterned WSe$_2$ where nanometer precision cuts were achieved using Helium Ion Microscopy (shown in Figure \ref{fig-HIM-WeSe2}b).

The cuts were found to enhance the localized intensity of the peak in the EEL spectra associated with the A exciton (regions shown in yellow). This highlights that the use of conventional STEM-EELS allows to add new spatial information to study how nanoscale variations in the specimen can affect the global excitonic signal. Furthermore, it showed that nanopatterning can be used to control local excitonic intensities.

While excitons in TMDCs are well studied in optical spectroscopy, q-EELS can provide information about their dispersion, and hence their propagation, their nature and type of excitation. q-EELS allows us to obtain data with high spatial, energy, and momentum resolution. Adding the information of transferred momentum opens up new opportunities to study optical modes in the electron microscope. However, a challenge is that excitons have a small cross-section in EELS, and with the quickly decaying intensity of EELS with increasing q, their detection becomes challenging and usually requires highly efficient detection capabilities such as direct detectors \cite{plotkin2020hybrid}. To-date only very few experimental studies have employed q-EELS to investigate excitons in TMDCs and other 2D materials, so this area of research is therefore still in its infancy. Existing studies employing q-EELS of excitons highlight different research questions around excitons~\cite{Habenicht_2018, Koitzsch2019NonlocalDF, Hong_2020, Hong2021, nerl2024mapping}. 
The q-EELS study by Suenaga \textit{et al.}\cite{Hong2021}  measured the energy-momentum dispersion of bulk excitons in  WSe$_2$ for small finite $q< 0.19 \text{AA} ^{-1}$. The bulk A exciton energy was found to vanish for $q>|\Gamma - K|/8$ or $q> 0.18 \text{AA} ^{-1}$. Importantly, a small shift in energy of the A exciton peaks was observed in the finite $q$ spectra ($\approx0.1$~eV) suggesting a parabolic dispersion behavior of the bulk exciton. 
In contrast, the studies by B{\"u}chner and Knupfer \textit{et al.}   ~\cite{Habenicht_2018,Koitzsch2019NonlocalDF} and Nerl \textit{et al}. \cite{nerl2024flat} covered a larger momentum transfer range to access phenomena outside of the light cone, including optically dark  excitons \cite{Malic2017DarkEI}. Inter-valley excitons are termed 'dark', \HCN{because their recombination requires a large momentum transfer between different valleys in the BZ, which cannot be provided by photons due to momentum conservation. This makes them invisible in standard optical spectroscopy.} They have instead been detected using angle-resolved photoemission spectroscopy (ARPES) in TMDCs including WSe$_2$ ~\cite{dong_2021}.  Habenicht \textit{et al. }\cite{Habenicht_2018,Koitzsch2019NonlocalDF} employed q-EELS of MoS$_2$ for high momentum transfer up to  $q= 1.33 \text{AA} ^{-1}$.  In their studies, the peaks in the high \textit{q} signal at the energy-loss positions of the A, B excitons were attributed to the possible  presence of dark excitons.  Nerl \textit{et al. } \cite{nerl2024flat} described an absence of defined features where dark excitons are expected, which is in contrast to the ARPES signatures obtained by Dong\textit{ et al.} \cite{dong_2021}. In contrast to the studies by Habenicht \textit{et al. }\cite{Habenicht_2018,Koitzsch2019NonlocalDF}, the intensity at high \textit{q} at the A exciton energy in the energy-momentum maps in Nerl \textit{et al}. \cite{nerl2024flat} was  attributed to the presence of exciton polaritons. When bulk excitons recombine, light is emitted. The bulk excitons can in turn couple to this emitted light to form an exciton polariton. At this point we would like to point to the review article by Basov \textit{et al. }\cite{basov2016polaritons} for more information on polaritons. The coupling of an exciton to any electromagnetic field is called exciton polariton and they are of great interest as they have recently been shown to undergo spontaneous coherence to form Bose Einstein (BE) condensates~\cite{penrosephysical, blatt1962bose, kasprzak2006bose, deng2010exciton, byrnes2014excitonpolariton, plumhof2014room}. This occurs when particles or quasiparticles which initially possess no phase relation, become coherent and form a single many-body wavepacket to condensate into the ground state of the system, once a specific parameter such as a threshold temperature has been reached. \HCN{Using electron-based techniques to investigate exciton polaritons has the advantage that scanning an electron beam across a specimen provides an incoherent source, thereby excluding the probe itself as a possible origin of coherence. While q-EELS cannot yet probe condensate populations directly, it can nevertheless provide insight into precursor stages of exciton–polariton formation. These stages are of particular interest as they constitute a necessary step toward BE condensation. It has also been shown that the nanomaterial itself can act as a natural microcavity, trapping photons through confinement at the material boundaries even in the absence of an external cavity \cite{khurgin2015two,munkhbat2019selfhybridized,talebi2022}. Accessing relative populations of electronic states using q-EELS may become feasible in the future as the technique continues to develop.}

\begin{figure}[h]
\centering
\includegraphics[width=1\textwidth]{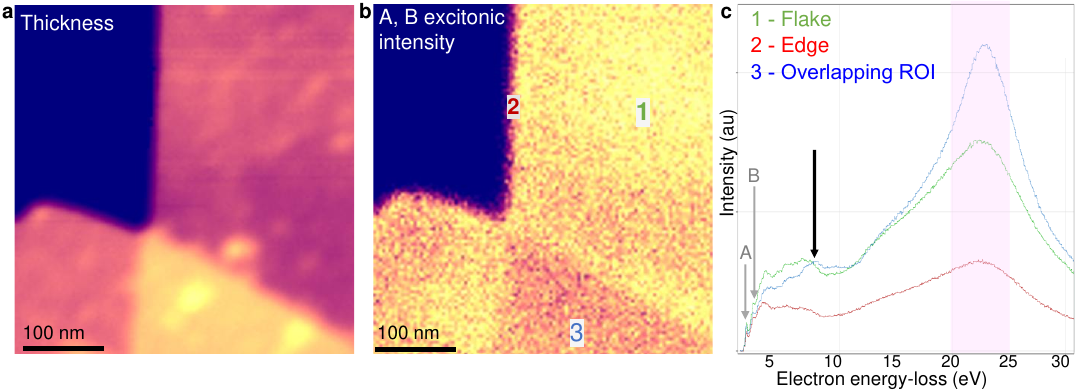}
\caption{The A,B excitonic intensities appeared suppressed in Moir\'e twisted homobilayers of WSe$_2$-WSe$_2$  using conventional STEM-EELS. \textbf{a}) the thickness map was extracted from the STEM-EELS map where the dark blue region shows the vacuum region, the pink area the regions of thin film WSe$_2$ and the yellow region shows the overlapping region of WSe$_2$. \textbf{b} The integrated intensity map of the A,B excitonic peaks from STEM-EELS shows that the region 3 is lower compared to the region 1. \textbf{c}) Overlaid EEL spectra from regions as denoted 1-3 in \textbf{b} shows that while there is a suppression of the A,B excitonic peak intensity \HCN{(grey arrows)}, there is a relative increase in the signal intensity of the higher energy peaks in the overlapping region between 7-10~eV which includes the $ \pi -\pi ^*$ plasmon peak of WSe$_2$ (black arrow). The $\pi +\sigma$ plasmon peak is denoted in the pink boxed area. (previously unpublished data)   }\label{fig-excitons-HomoBilayer_WSe2}
\end{figure}

The above suggests that in the electron microscope we might be probing an overlapping signal of both, bulk excitons and exciton polaritons.  The energy-momentum maps in Nerl \textit{et al.}  \cite{nerl2024flat} (shown here in Figure \ref{fig-excitons-HomoBilayer_WSe2},c) showed a signal at the A and B excitonic energy that persisted right across the BZ. Several possible sources contributing to the signal at high \textit{q} were mentioned  including the possible presence of excitonic polarons which can form when excitons interact with acoustic or optical phonons via coupling to the deformation potentials
\cite{thilagam2015excitonic}. To investigate the possibility that phonons could contribute momentum via dual scattering processes, cryo q-EELS was employed to suppress the electron energy-gain through phonons \cite{PhysRevB.106.195431, minson2023quantitative}. It was speculated that since no detectable changes were observed, phonons could only contribute in a minor way at finite \textit{q}. Using PL, Huang \textit{et al.} \cite{huang2016probing} found that the phonons contributed, leading to a shift with decreasing temperature in the energy of the A exciton peak in WSe$_2$ which is consistent with phonon absorption being the sole phonon contributor at low temperature and phonon emission and absorption contributing at room temperature. Probing the involvement could be approached in the future by using careful comparative studies where parameters such as temperature could be adjusted to tune the phonon population in a targeted manner. 
\HCN{In addition to the phenomena described above,} another research focus is interlayer excitons, where the exciton is formed by an electron and a hole from two separate layers. The exciton binding energy is thereby modulated via coupling to the other layer. Since the local atomic arrangement changes dramatically with twist, the properties of the TMDC heterostructures are tunable by the twist angle between the layers which provides a highly tunable platform for strongly correlated electron physics \cite{devakul2021magic}. \HCN{While interlayer excitons are of great interest in twisted and heterobilayer TMDCs due to their long lifetimes and spatially indirect nature, they have not yet been unambiguously observed in EELS experiments. This is likely due to their comparatively small inelastic scattering cross-section relative to intralayer excitons, which makes their direct detection by electron beams extremely challenging. To date, evidence for interlayer excitons has been provided mainly through optical spectroscopies, which have revealed long-lived excitons in MoSe$_2$/WSe$_2$ heterobilayers~\cite{rivera2016valley}, demonstrated polarization switching and electrical control of excitonic valley pseudospin~\cite{ciarrocchi2018electrical}, and uncovered Moir\'e-trapped interlayer excitons~\cite{jin2019observation,rivera2018interlayer}. In contrast to interlayer excitons, which remain elusive in EELS, conventional STEM-EELS studies have instead focused on how twist angle modifies the intralayer (bulk-like) excitons of the constituent layers in heterostructures \cite{gogoi2019layer,susarla2022hyperspectral, woo+23prb}. }
In addition, the bulk excitons of the individual components of a heterostructure are dependent on the twist angle when the layers are stacked. For example, when homobilayers (two separate monolayers) of MoS$_2$ and WSe$_2$ are rotationally aligned (or anti-aligned), the A, B excitons become considerably suppressed compared to the bulk excitons in homobilayers with other twist angles as shown in Gogoi \textit{et al.}~\cite{gogoi2019layer} and illustrated with previously unpublished data in Figure \ref{fig-excitons-HomoBilayer_WSe2}a-c. In the STEM-EELS map in Figure \ref{fig-excitons-HomoBilayer_WSe2}a, the thickness was mapped for a region of two overlapping thin films of WSe$_2$ to form a homobilayer of WSe2/WSe$_2$.  The corresponding A, B excitonic intensity maps from STEM-EELS show that the A, B intensities are suppressed in the overlapping regions compared to the edge and thin film reference regions. When overlaying the spectra from the thin film (denoted with 1), edge region (denoted with 2) and overlapping region (denoted with 3) shown in Figure \ref{fig-excitons-HomoBilayer_WSe2}b-c, it is apparent that the most prominent energy peak in the energy-loss region of the C excitonic transitions is found at higher energy (arrow).  It has been suggested previously that the C peak excitonic transitions are composed of a multi-feature peak due to their diverse $k$-space origin of the different C transitions \cite{Hong2021}. Hence what appears to be a change in the energy position of the major C peak might also be caused by a spectral weight redistribution in the STEM-EELS in the overlapping region labeled '3' in Figure \ref{fig-excitons-HomoBilayer_WSe2}b. \HCN{This interpretation is consistent with the findings of Susarla \textit{et al.}, who used conventional STEM-EELS mapping of heterobilayers to directly visualize exciton localization in Moir\'e heterostructures, providing experimental evidence that Moir\'e patterns modulate excitonic properties at the nanometer scale \cite{susarla2021mapping}.} 
 
\begin{figure}[h]
\centering
\includegraphics[width=1\textwidth]{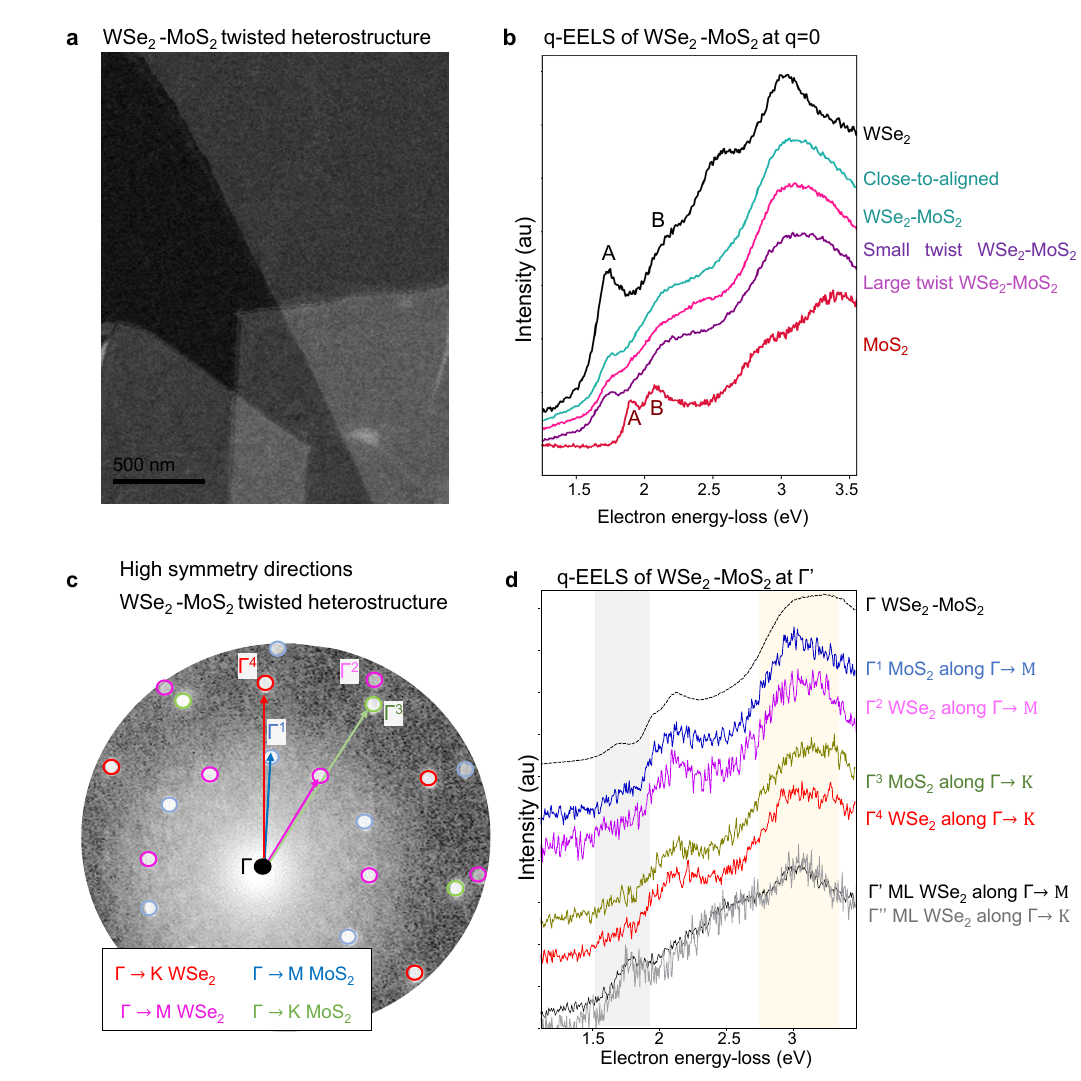}
\caption{ Heterostructure of twisted WSe$_2$-MoS$_2$ shows twist-dependent variations in the spectral peaks in the excitonic loss region using q-EELS. \textbf{a}) STEM image of the heterostructure. \textbf{b}) q-EELS at q=0 for WSe$_2$ (black), close-to-aligned WSe$_2$-MoS$_2$ (turquoise), small twist WSe$_2$-MoS$_2$ (purple), large twist WSe$_2$-MoS$_2$ (pink) and MoS$_2$ (red). It shows the suppression of the A,B excitons in the Moir\'e heterostructures. \textbf{c}) Schematic of the diffraction space of the large twist WSe$_2$-MoS$_2$ showing how the overlapping directions of high symmetry increase complexity of analysis. d) Overlaid (vertically offset) spectra from q=0 at $\Gamma$ (black), $\Gamma ^1$ along $\Gamma \rightarrow M$ of MoS$_2$ (blue), $\Gamma ^2$ along $\Gamma \rightarrow M$ of WSe$_2$ (pink),  $\Gamma ^3$ along $\Gamma \rightarrow K$ of MoS$_2$ (green), $\Gamma ^4$ along $\Gamma \rightarrow K$ of WSe$_2$ (red),  $\Gamma '$ along $\Gamma \rightarrow M$ of WSe$_2$ (black), $\Gamma ''$ along $\Gamma \rightarrow K$ of WSe$_2$ (grey). Again the A,B excitons appear suppressed in all of the heterostructure spectra (boxed green region of loss spectra). There is some unexpected spectral variations in the C excitonic loss region depending on the offset $\Gamma$ points chosen (yellow boxed region). }\label{fig-excitons-hetero}
\end{figure}

Furthermore, to retain the transferred momentum information of excitonic signals in heterostructures of TMDCS, we analyzed q-EELS of one of the most common TMDC heterostructures, WSe$_2$-MoS$_2$ (see HAADF-STEM image in Figure \ref{fig-excitons-hetero}a). Figure \ref{fig-excitons-hetero}b compares q-EELS at zero momentum transfer for samples with close-to-aligned, small twist, large twist and anti-aligned heterostructures of WSe$_2$-MoS$_2$. Like for the STEM-EELS maps, the A and B excitonic peak intensities in the spectra from $q=0$ show significant variations depending on inter-layer twist, but they are suppressed for all the non-zero twist angles, which is in agreement with previous STEM-EELS studies \cite{gogoi2019layer,susarla2022hyperspectral, woo+23prb}. In fact, in the case of the largest twist angle, the intermediate excitonic peak does not coincide with excitons of either individual layers.  The schematic in Figure \ref{fig-excitons-hetero},c shows how the complexity increases rapidly once two layers of distinct TMDCs are investigated using q-EELS.  The two near anti-aligned layers in a heterostructure of MoS$_2$ and WSe$_2$, have the high symmetry direction $\Gamma \rightarrow M$ of MoS$_2$ overlap with the high symmetry direction $\Gamma\rightarrow K$ from WSe$_2$ and vice versa. Nevertheless the q-EEL spectra from specific diffracted spots can be analyzed to compare for difference. The q-EEL spectra in Figure \ref{fig-excitons-hetero}d show that the A exciton is suppressed in the heterostructure at the $ \Gamma$ point as well as at $ \Gamma^{1}$,  $ \Gamma^{2}$, $ \Gamma^{3}$and $ \Gamma^{4}$ compared to the q-EELS at  the  $ \Gamma'$ and the  $ \Gamma''$ points of the individual monolayers of WSe$_2$. In addition, we observe a variation in the higher energy excitonic peaks when comparing the q-EELS along $\Gamma \rightarrow M$ of MoS$_2$ and WSe$_2$ as well as along $\Gamma \rightarrow K$ of MoS$_2$ and WSe$_2$ (yellow box). This most likely reflects the difference in energy-momentum dispersion behavior of the higher energy excitonic peaks \cite{Hong2021} and/or difference in sampling of the \textit{q} space due to the changes in the scattering geometry of the set-up. 

In heterostructures with a smaller relative twist it becomes difficult to extract the q-EELS from closely oriented diffracted spots of the heterostructure. In addition, signal-to-noise is increasingly challenging as monolayers and thin films have a small scattering cross section leading to the intensity at diffracted spots being very low. It is worth noting that we observed an absence of defined features at finite \textit{q} in the regions where dark excitons are expected. This is surprising considering that dark excitons have been detected along  $\Gamma \rightarrow K$ using ARPES but it is in agreement with an earlier study by some of the authors \cite{nerl2024flat}. The discrepancies in research findings might be due to a difference between techniques which should be investigated in the future. 

Employing q-EELS also opens up the opportunity to study excitons at high momentum transfer that lie outside of the light cone and remain therefore inaccessible to optical techniques, for instance in hBN. Its honeycomb structure based on \textit{sp}\textsuperscript{2} covalent bonds leads to its atomically flat surface and interesting excitonic signature which make hBN an attractive candidate for a range of applications \cite{caldwell2019photonics}. Nevertheless, there remained some controversy regarding whether or not hBN is in fact an indirect semiconductor \cite{cassabois2016hexagonal}. The relative high energy of the lowest-energy exciton in hBN combined with what has been shown to be a complex anisotropic dispersion behavior at high momentum transfer using non-resonant inelastic x-ray scattering (NRIXS) and theoretical calculations by  Galambosi \textit{et al.}~\cite{gala+11prb}, make hBN excitons particularly challenging to study.  The nature of the hBN excitonic effects have previously been identified using BSE calculations~\cite{arnaud2006, wirtz2006,fugallo+15prb, Aggoune2018DimensionalityOE}. \HCN{Both experimental optical studies and first-principle calculations have shown that the spectral weight redistributes at the M' points along $\Gamma \rightarrow K$ due to the directional dependence of oscillator strengths with increasing momentum \cite{gala+11prb, fugallo+15prb}. This redistribution has been linked to strong excitonic effects at $\approx 7\mathrm{eV}$, where $\pi -  \pi*$  transitions give rise to a pronounced peak in the joint density of states (JDOS).}  The excitonic anisotropy was confirmed using q-EELS \cite{foss+17prb, schuster2018direct} and then later using energy-momentum q-EELS mapping across the entire BZ \cite{nerl2024mapping}. The mapping results from q-EELS \cite{nerl2024mapping}  are presented in Figure \ref{fig-hBN}b-c where the relative spectral intensity changes significantly along $\Gamma \rightarrow K$, especially  at the M' point of the BZ. This was not the case for the q-EELS along  along $\Gamma \rightarrow M$ (not shown here, cf.  reference \cite{nerl2024mapping}). The results from q-EELS show the complex anisotropic dispersion behavior at high momentum transfer. 
Selected individual q-EEL spectra showing the spectral intensity redistribution of the excitonic intensities at the $M'$ point at $|q| = 2.5 \text{AA} ^{-1}$ compared to $q=0$ and at the $K$ point along $\Gamma \rightarrow K$ are presented in Figure \ref{fig-hBN},d. These are in agreement with results for the calculated dynamic structure factor for \textit{q} along $\Gamma \rightarrow K$ from the solution of the Bethe-Salpeter equation (BSE) \cite{gala+11prb} as well as NRIXS data \cite{fugallo+15prb} which are also presented in in Figure \ref{fig-hBN},e. The calculated dynamic structure factor for \textit{q} along $\Gamma \rightarrow K$ from GW-RPA, \HCN{a perturbative approach based on single-particle Green's function $G$ and the screened Coulomb interaction $W$ within the random phase approximation,} is also shown.  This shows that q-EELS in modern microscopes can rival the results from non-resonant x-ray techniques in some cases. 
\clearpage

\subsection{Phonons}\label{sec-phonons}
Quantized lattice vibrations, known as phonons, play an important role in material properties including thermal conductivity, phase transformations, and phonon-driven superconductivity \cite{snyder_complex_2008, pernot_precise_2010, biswas_high-performance_2012, maldovan_sound_2013, cao_unconventional_2018, xie_brief_2023}. Studying phonons, including their localized modes, is crucial to gain fundamental understanding and control with the aim to ultimately optimize material and device performance in novel nano- or quantum electronic devices. Phonon propagation is described by their dispersion relation, which are in turn affected by geometry, e.g., through dimensionality and local atomic structure \cite{hage_phonon_2019, haas2023atomic, yan_single-defect_2021, yan_nanoscale_2024}. In general, phonons are delocalized. However, a reduction in dimensionality at the level of individual atoms, interfaces or grain boundaries will typically impose a local character. Probing such localized phenomena with STEM-EELS requires sufficient collection of signal from associated localized scattering events \cite{dwyer_localization_2014, dwyer_2016}.

As the amount of energy transferred to phonons is very small (in the meV range), phonons are often probed by means of optical techniques, such as Fourier-transform infrared absorption~\cite{griffiths1983fourier,griffiths2007fourier} or Raman spectroscopy~\cite{cardona1982light}, both of which are limited by the momentum that can be transferred to the material excitation and by symmetry-inherent selection rules~\cite{loudon1964theory}.  Improvements in the energy resolution of STEM instruments a decade ago have opened up a whole new information level, allowing access to phonon excitations in materials using conventional STEM-EELS and q-EELS measurements \cite{krivanek_vibrational_2014, miyata2014measurement}.
In crystalline materials  both, vibrational bulk modes and surface modes can be probed using STEM-EELS. This was shown by Lagos \textit{et al.} \cite{lagos2022advances} by mapping the surface modes excited on the corners and faces of a single MgO nanocube, as well as individual bulk acoustic and optical modes. Vibrational STEM-EELS has also been shown to provide isotope sensitivity \cite{hachtel_isotopes_2022, senga_isotope_2022} and allow for the detection of vibrational modes of functional groups of organic molecules \cite{rez_bio_2016}. In addition, vibrational STEM-EELS allows for highly local temperature measurements when considering both the gain and loss parts of the spectrum, using the principle of detailed balance \cite{PhysRevLett.17.379, lagos_thermometry_2018, idrobo_temperature_2018}. 

Scattering of vibrational modes by fast electrons is typically discussed in terms of  long-range 'dipole scattering' and localized 'impact scattering' \cite{ibach_mills_1982, egerton_2011}. While impact scattering has a broad angular distribution, dipole scattering is peaked in the forward scattering direction. When permitted, dipole scattering can be expected to dominate the vibrational EEL spectrum at small scattering angles \cite{ibach_electron_1982}. For non-polar materials such as silicon and graphene only negligible dipole scattering can be expected since lattice oscillations are not associated with a dipole moment. For polar materials (e.g., boron nitride) however, dipole scattering is expected due to non-zero dipole momenta of lattice oscillations. The latter meaning that dark field and bright field geometries can be used to select for localized and delocalized phonon scattering \cite{dwyer_localization_2014}. 

Taking advantage of the above, atomically resolved phonon mapping was first demonstrated by Hage \textit{et al.} \cite{hage_phonon_2019} and Venkatram \textit{et al.} \cite{venkatraman_vibrational_2019}. Hage \textit{et al. } used an off-axis beam geometry to significantly reduce the dipole contribution in spectra from hBN, favoring the localized vibrational scattering and thereby following the approach by Dwyer \textit{et al}. \cite{dwyer_2016}. In contrast, Venkatraman \textit{et al.} employed an on-axis aperture to detect localized impact scattering in non-polar silicon. Thereafter, the vibrational signature of a single silicon defect in graphene was reported by Hage \textit{et al.} \cite{hage_single-atom_2020} . Later in the same system, bonding sensitivity of silicon defects in two different configurations was discerned by Xu \textit{et al}. \cite{xu_single-atom_2023}. The achievements above show that experimental geometry (i.e., the combination of convergence and collection angles) plays a crucial role in optimizing vibrational STEM-EELS experiments. The obtainable extreme spatial resolution of vibrational EELS has been demonstrated in studies of the atomic scale of grain boundaries \cite{haas2023atomic, hoglund_direct_2023, yan_nanoscale_2024}, single, extended defects \cite{yan_single-defect_2021}, and interfaces \cite{shi_interface_2024, qi2021measuring}, see review of Haas \textit{et al.} \cite{haas_perspective_2024} for an in-depth survey on atomically resolved vibrational EELS. 

\begin{figure}[htbp]
\centering
\includegraphics[width=0.9\textwidth]{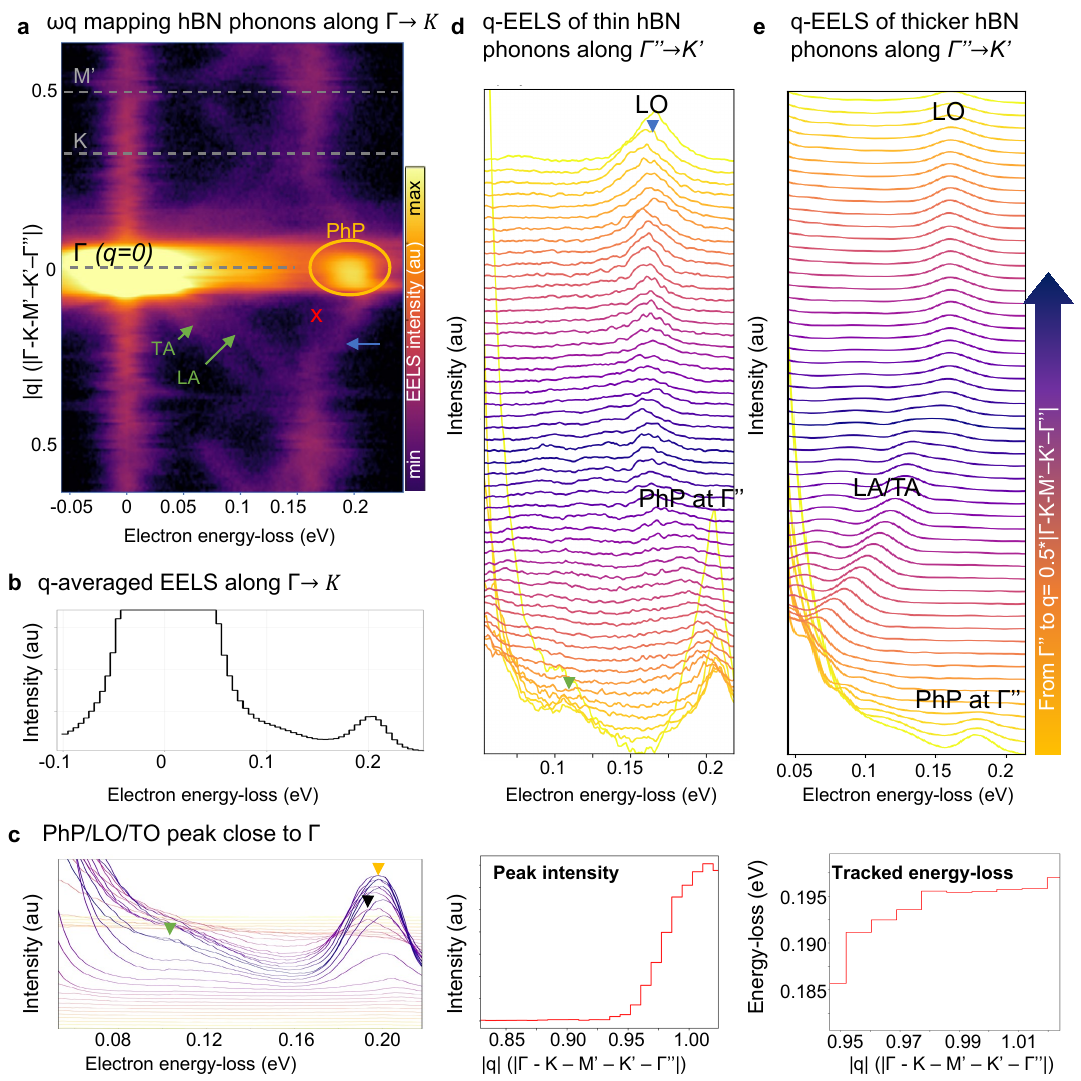}
 \caption{a) Energy momentum ($\omega q$) map in log scale of thin hBN phonons acquired using a slit aperture placed to select along $\Gamma \rightarrow K$, where the additional information provided through q-EELS allows to distinguish between polaritonic and bulk phonon modes. The transverse acoustic (TA) and longitudinal acoustic (LA) modes are denoted by green arrows, while the longitudinal optical (LO) mode is denoted by a blue arrow. The red 'x' marks the region where the transverse optical (TO) is observed. The polaritonic signal (PhP) is dominant at $q \rightarrow 0$ (yellow circle) and masks other signals in the summed spectra shown in b). This is comparable to the masking that occurs in conventional STEM-EELS where the momentum information is averaged.  c) q-EEL spectra show the PhP (orange), LA/TA (green) and LO/TO (black) peaks close to $\Gamma $. When tracked, the major peak contains a mixture of PhP contributions and LO/TO contributions for small \textit{q}. As expected, the peak intensity drops dramatically with increasing \textit{q}. The energy-loss peak maximum shifts at small q, from $0.195\,\mathrm{eV}$ at $\Gamma $ to $0.186\,\mathrm{eV}$ at $0.95|q|$. This could indicate an actual peak shift or a change in the relative peak intensity of PhP, and LO/TO peaks.  d) Individual q-EEL spectra were chosen starting from $\Gamma ''$ along $\Gamma ''\rightarrow K'$ to avoid the contributions from the central undiffracted beam. They show the evolution of the peaks across the BZ, including the polaritonic modes (annotated with PhP) in few-layer thin hBN (approx.~3-layer) and e) thicker hBN ($\approx15\,\mathrm{nm}$). Please note that in the thin hBN, the LA/TA modes are only visible in the $\omega q $  map (log scale) and not the individual spectra presented here. This is due to the scaling of the spectra as the LO modes are significantly stronger in intensity. This is not the case for the thicker hBN, where LA/TA as well as the LO modes are well visible at finite momentum transfer spectra.   }\label{fig-phonons-hBN}
\end{figure}

Sacrificing part of the attainable spatial resolution, nanoscale momentum-resolved EELS opens the opportunity to measure phonon dispersion relations while retaining a high degree of spatial selectivity to investigate both, the \textit{q-}dependency of phonon modes and their local behavior. The acquisition of phonon dispersion curves using serial q-EELS was first demonstrated by Hage \textit{et al.} \cite{hage_nanoscale_2018} to distinguish different phonon branches of the two allotropes hBN and cubic boron nitride (cBN) as a function of transferred momentum. Next, Senga \textit{et al.} \cite{senga_position_2019} measured phonon dispersion curves of graphene, h-BN and graphite across several BZs and along different high symmetry lines using serial q-EELS. They found the absence of optical modes in spectra for $q \rightarrow 0$ in graphene and graphite which they attributed to perfect screening by the valence density. Hence, in the optical limit, only the longitudinal acoustic (LA) mode is excited in graphene and graphite. However, the higher momentum electrons (exceeding the first BZ) can excite optical phonons in graphene and graphite as the valence screening is insufficient for finite \textit{q} \cite{senga_position_2019}. In contrast, in polar material such as hBN, both LA and longitudinal optical (LO) modes \HCN{couple to the electron beam even at small momentum transfers~\cite{froehlich1954electrons}, and thus remain detectable already within the 1st BZ and beyond (see Figure \ref{fig-phonons-hBN})}.
Additionally, spatial phonon mapping, where the collection aperture is displaced away from the optical axis, allows to detect localized modes, including enhancement of acoustic phonon intensity at edges and structural defects. By contrast, the optical phonon signal is mainly sensitive to specimen thickness~\cite{senga_position_2019}.

The introduction of direct electron detectors in 2019 allowed for the simultaneous recording of energy and momentum to generate vibrational $\omega q $ maps rather than individual spectra, \CE{first demonstrated using hBN \cite{bleloch_advances_2019, plotkin2020hybrid}.} This is made possible by their superior dynamic range and detection efficiency compared to conventional charge-coupled device (CCD) detectors, both of which significantly reduced experimental operation time associated with serial q-EELS. Specifically, the increase in the dynamic range of the direct electron detectors allows for the high intensity ZLP and the lower intensity low loss signals to be covered within the same acquisition which has a number of benefits for data acquisition and accuracy of interpretation. Recently, Li \textit{et al.} \cite{li_systematic_2024} studied the phonon dispersion of graphene beyond the first BZ using $\omega q$ mapping. They observed a systematic absence of optical phonon signals in certain higher-order BZs. This absences was attributed to destructive interference between inelastically scattered electrons, so that the optical phonon loss peaks appear only at the $\Gamma$ points of selected BZs. \HCN{Importantly, this effect reflects a “missing intensity” in the $\omega q$ maps, the phonon branches themselves still exist, but their spectral weight vanishes from the EELS signal due to interference.}

Phonon polaritons (PhPs) are excited in polar materials when optical phonons having an oscillating dipole moment couple to electromagnetic waves induced by the beam electron \cite{xu_one-dimensional_2014}, i.e., through dipole scattering. Their energy relative to the bulk TO/LO modes is small, and PhPs were first resolved as a separate peak from the bulk phonon signal using STEM-EELS by Batson and Lagos \cite{batson2017characterization}. Hexagonal BN has been shown to sustain hyperbolic PhPs due to its anisotropy and permits momenta beyond the light cone, with the property of high light confinement investigated for applications such as nanosensing and quantum optics \cite{basov2016polaritons, li_direct_2021,jacob_hyperbolic_2014}. Govyadinov \textit{et al.} \cite{govyadinov_probing_2017} employed STEM-EELS to confirm the identity and map the propagation of hyperbolic PhPs in thin film hBN. The PhPs were found to propagate along the edge surfaces as guided modes, due to them only being excited at surfaces parallel to the optical axis, i.e. at flake edges where the in-plane anisotropy is strong. These findings are consistent with earlier works using scanning near field optical microscopy (SNOM) \cite{xu_one-dimensional_2014}.

As dipole scattering and consequently the excitation of PhPs dominates in polar materials such as hBN, the signal arising from PhPs will typically mask the bulk phonon signal in STEM-EELS at $q=0$ and low $q$.
Using $(\omega q)$-mapping the dominating PhP signal in hBN is indicated in Figure \ref{fig-phonons-hBN}a, where the PhP signal is circled in orange and the LO mode at finite \textit{q} is marked with the blue arrow. Figure \ref{fig-phonons-hBN}b shows a q-EEL spectrum integrated along \textit{q}, comparable to the response signal typically collected by a conventional STEM-EELS geometry. Q-EELS therefore allows for mapping the low intensity phonon modes at finite \textit{q} that originates from impact scattering, and for separating the PhPs from bulk phonon signals. Figure \ref{fig-phonons-hBN}c shows the PhP and/or LO/TO modes close to the $\Gamma$ point, with the PhP (orange), LA/TA (green) and LO/TO (black) peaks. The major peak appears to contain a mixture of PhP contributions and LO/TO contributions for small \textit{q}  which are challenging to discern. The intensity of the main peaks falls drastically with increasing momentum. The tracked maximum shifts from above $0.195\,\mathrm{eV}$ at the $\Gamma$ point to $0.1886\,\mathrm{eV}$ at $|q|=0.95*|\Gamma KM'K'\Gamma''|$, which represents the shift separating PhPs and LO/TO peaks \cite{batson2017characterization}. Figure \ref{fig-phonons-hBN}d shows individual q-EEL spectra of thin h-BN, and \ref{fig-phonons-hBN}e from thicker hBN. The q-EELS  were chosen to start from $\Gamma ''$ along $\Gamma ''\rightarrow K'$ to avoid the overpowering contributions from the central undiffracted beam. In the thin hBN, the LA/TA signal is better discernible in the $\omega q $ map in Figure \ref{fig-phonons-hBN}a displayed in log scale, and not in the individual spectra due to their low relative intensity. In thicker hBN, the LA/TA peaks have much stronger signal intensity than the LO/TO modes at low \textit{q}.

Nonanalytic behavior of optical phonons was observed in monolayer hBN suspended on polycrystalline copper foil by Li\textit{ et al. } \cite{li_observation_2024} using REELS. The dispersion of the LO modes in this case was found to follow a v-shaped behavior close to the $ \Gamma $ point. While the TO and LO modes are degenerate at $q \rightarrow 0$, the peaks split for non-zero small \textit{q}. This splitting has been shown beautifully by Li \textit{et al. }\cite{li_observation_2024}.

The following examples highlight how vibrational q-EELS can be combined with or complemented by spatially resolved EELS. Qi \textit{et al.} \cite{qi_four-dimensional_2021} measured phonon dispersions across a single BN nanotube by combining spatial scanning with $\omega q$-mapping, such that a $\omega q$-map is acquired at each beam position \cite{Wu2023FourdimensionalEE}. This approach revealed how the relative contribution from different phonon peaks evolve with increasing momentum transfer at specific locations. Consistent with findings on other defects and interfaces \cite{yan_single-defect_2021, haas2023atomic, yan_nanoscale_2024, senga_position_2019}, they observed that acoustic modes are highly sensitive to structural defects, whereas optical modes remain largely unaffected. Applying the same method to the diamond/cubic-BN  interface~\cite{qi2021measuring}, they identified vibrational response specific to the interface. Similarly, Yan \textit{et al.} \cite{yan_single-defect_2021} combined serial q-EELS with spatial vibrational mapping of a single stacking fault in SiC. They detected a significant redshift of the transverse acoustic mode localized to within a few nanometers of the stacking fault, while the optical modes remained unaffected. These studies demonstrate the intrinsic trade-off between momentum and spatial resolution in vibrational q-EELS, which limits the information available simultaneously.

\HCN{Beyond spatially resolved mapping, vibrational q-EELS can also be tailored to extract directional information on the phonon propagation. Gadre \textit{et al.} \cite{gadre_nanoscale_2022} implemented a momentum-difference setup to recover the directional momentum dependence of the phonon signal.} This allowed them to study phonon dynamics at abrupt and gradual Si interfaces of a silicon–germanium quantum dot. Briefly, the energy-loss was recorded at opposite positions in momentum space with respect to the central beam, and subtracted to reveal the net momenta. In this way, the so-called "phonon flux" is obtained by correlating the net momenta with the phonon group velocity, to observe how the flux is affected by the abrupt and gradual interface. Enhanced phonon intensity was observed at the abrupt interface, which they attribute to strong reflection arising from sudden changes in the phonon density of states, effectively hindering propagation such that thermal conductivity is limited.

\section{Complexities, Challenges, and Outlook}\label{sec-Outlook}

\HCN{Recent advances in q-EELS now make it possible to probe the fundamental excitations of 2D materials with a breadth and level of detail that was long thought out of reach. Owing to its unrivaled ability to simultaneously access high energy, momentum, and spatial information, the technique has opened new opportunities for exploring collective excitations in low-dimensional systems. Despite these successes in probing plasmons (see Section \ref{sec3-plasmons}), excitons (Section \ref{sec-excitons}), phonons (Section \ref{sec-phonons}) and their coupling in 2D materials, q-EELS still faces significant hurdles. Establishing q-EELS as a broadly applicable methodology will require addressing both technical challenges and fundamental constraints . The following subsections highlight emerging research frontiers, the need for experimental innovations such as cryogenics, tomographic mapping, and \textit{in situ} methods, challenges of quantification and benchmarking, as well as advances in theory and machine-learning–based analysis. }\\

\HCN{\textbf{Emerging frontiers}. Most recently, q-EELS has been employed to study magnons in materials, and the only published experimental evidence to date has been by Kepaptsoglou \textit{et al.}~\cite{kepaptsoglou2024magnon}. This builds on earlier theoretical predictions on magnons by Wu \textit{et al.}~\cite{wu2018magnon}, and subsequent work on temperature diffuse scattering of magnons by Castellanos \textit{et al.}~\cite{castellanos2023unveiling}. In 2024, the same group presented a method for calculating the angle-resolved electron energy-loss spectra resulting from phonon or magnon excitations~\cite{castellanos2025dynamical}. These studies highlight the difficulty of  separating phonon from magnon diffuse scattering, since magnons are several orders of magnitude weaker yet appear at similar energy losses.}
\HCN{Similar challenges arise in the phonon domain, which remains another frontier for q-EELS. In particular, TMDCs and other low-dimensional materials host phonon modes at much lower energies than hBN and graphene. This makes it especially challenging to detect and analyze them~\cite{cai2014mos2phonons,gadre_nanoscale_2022}, pushing the limits of energy resolution. Multiphonon excitations remain an even more ambitious goal, with no direct q-EELS measurements reported to date~\cite{barthel2024phonon, castellanos2025dynamical}. Furthermore, multiple scattering processes must be carefully accounted for to disentangle phonon contributions, a requirement that complicates assignment and quantitative analysis of spectra. In addition, recent theoretical advances show that phonon dispersions measured with q-EELS are governed by eigenvector selection rules and coherent interference effects that can suppress or unfold entire branches. Pfeifer \textit{et al.}~\cite{pfeifer2025phonon} introduced the concept of an 'interferometric Brillouin zone' to capture these effects and proposed efficient strategies for simulating phonon spectra.}\\

\HCN{\textbf{Experimental innovations.}  Advances in both magnons and phonons are likely to depend on experimental innovations, especially the adoption of cryogenic operation. Cooling not only sharpens vibrational features but also provides direct access to temperature-driven phase transitions, collective quantum phenomena, and potentially superconductivity~\cite{yan_single-defect_2021}. Liquid-helium stages are now commercially available and have been used for high-resolution imaging~\cite{mun2024lhe} and are under development for EELS~\cite{johnson2024uheels}. To our knowledge no published q-EELS vibrational mapping at liquid-helium temperature has yet been reported.}

\HCN{Heterostructures of 2D materials introduce yet another element of complexity. Using q-EELS to probe hybrid and coupled modes across the BZ is still at an early stage, and the twist dependence of properties effectively adds another dimension to the dataset. Each twist angle requires preparation and characterization of a new specimen, and the resulting complex diffraction patterns ideally demand mapping of the full reciprocal space rather than just high-symmetry lines. This is extremely time-consuming with slit or circular apertures and realistically calls for automation. Automated tomographic acquisitions of the reciprocal space using slit apertures, analogous to existing tomographic reconstruction approaches in electron microscopy \cite{midgley2009tomography}, could generate comprehensive maps of the diffraction space. Such datasets would provide access to anisotropies and emergent features that cannot be captured when restricting measurements to high-symmetry directions. Although challenging due to long acquisition times and demanding data analysis of large data volumes, such approaches may reveal anisotropy in polariton dispersions and enable direct visualization of emergent properties such as Moir\'e minibands~\cite{jin2019observation} or twist-angle–dependent hybridization~\cite{seyler2019signatures}.}

\HCN{The scope of q-EELS continues to broaden in parallel with the wider 2D materials field, now spanning magnetic~\cite{burch2018magnetism}, thermoelectric~\cite{snyder_complex_2008}, superconducting, and exotic topological phases~\cite{manzeli20172d, hasan2010colloquium,qian2014quantum}. The emergence of intrinsic 2D magnetism in materials such as CrI$_3$ and Fe$_3$GeTe$_2$~\cite{gong2017discovery} opens new opportunities to probe magnon dispersion, spin dynamics, and phase transitions with q-EELS, provided energy resolution and signal levels can be sufficiently improved. Realizing this potential will require experiments that capture dynamics under external control. \textit{In situ} capabilities are therefore becoming increasingly important. Combining q-EELS with external electrical bias, gating, or mechanical strain promises real-time mapping of evolving excitations in excitonic, plasmonic, and phononic behavior. Barantani \textit{et al.}~\cite{barantani2025trqEELS} recently demonstrated the first ultrafast pump--probe q-EELS experiments revealing the momentum-resolved interplay of plasmons and phonons in graphite on femtosecond timescales. Another promising application of \textit{in situ} q-EELS is carrier-density tuning in graphene/hBN stacks, which may allow continuous observation of hybrid plasmon-phonon polariton formation and decay. Such behavior has previously been demonstrated with SNOM in the landmark work of low-loss, highly confined plasmons in 2D heterostructures  by Woessner \textit{et al.}~\cite{woessner2015highly}, but not yet achieved with q-EELS.}\\

\HCN{\textbf{Quantification and benchmarking.} A further longstanding challenge concerns the quantification of spectral intensities, especially since in q-EELS this extends to spectral intensities across the BZ. Without robust theoretical underpinnings, only relative intensities are typically compared. Leon \textit{et al.}~\cite{leon_unraveling_2024} demonstrated one promising approach by extracting the q-dependent loss function from experimental q-EELS using normalization to the \textit{f}-sum rule, thereby aligning experimental intensities with simulated loss functions. Extending such quantification methods to measure absolute intensities would enable future studies of relative population sizes of excitons, critical for investigating collective phenomena like Bose-Einstein condensation. }

\HCN{Finally, the broader impact of q-EELS will depend on developing robust experimental protocols (to minimize damage and artifacts), multi-modal combinations (with SNOM, STM, or X-ray spectroscopies), and quantitative frameworks for benchmarking against established techniques~\cite{garciadeabajo2010optical,krivanek_vibrational_2014}. In particular, comparisons with inelastic neutron and X-ray scattering for phonons and magnons, and Raman spectroscopy for excitons and phonons~\cite{baron2015elastic}, will be essential for validating q-EELS and establishing its unique strengths. While inelastic neutron and X-ray scattering provide bulk-averaged dispersions from large crystals, and Raman offers optical access near $q\approx 0$, q-EELS delivers complementary information by resolving excitations with nanometer spatial selectivity and full momentum control in thin, electron-transparent samples. Addressing these challenges will significantly deepen our understanding of 2D materials and open new technological possibilities.}\\

\HCN{\textbf{Theory and data analysis.} Theoretical progress requires moving beyond the local dielectric approximation, which is particularly important for ultrathin 2D systems. Incorporating spatial dispersion, i.e. $\epsilon(\omega,q)$, captures nonlocal screening and anisotropy that strongly affect excitons and polaritons at finite momentum. Modern \textit{ab initio} and many-body computational approaches (DFT, GW, BSE, TDDFT) now provide  detailed predictions of exciton fine structure, phonon anomalies, and hybrid modes. q-EELS is uniquely positioned to validate these predictions over an extended $q$-range~\cite{cudazzo2011dielectric, caruso2015band, despoja2013plasmon, trolle2017model, qiu2013manybody, ugeda2014giant, cocchi2015electronic}. Together these advances are key to quantifying many-body interactions, understanding damping, and moving beyond simple additive stacking models in van der Waals heterostructures. }

\HCN{As datasets grow larger and more complex, automated analysis is becoming indispensable. The analytical challenge posed by multi-dimensional q-EELS data (real space, energy, momentum) and the prospective multi-modal experiments make integration with artificial intelligence (AI) and machine learning (ML) essential~\cite{botifoll2022machine}. Both unsupervised and supervised learning methods, including deep and manifold learning (e.g. UMAP, t-SNE), can automate feature extraction, highlight subtle spectral trends, and guide experimental optimization. Recent advances demonstrate the potential of autoencoder-based denoising for rapid, low-signal EELS acquisition~\cite{pate2021}, clustering for robust pattern discovery in large spectral datasets~\cite{blancoportals2022}, and supervised neural networks for reliable elemental and oxidation state identification~\cite{delpozo2023}. At the same time, careful attention to pre-processing and data quality remains crucial, while emerging anomaly-detection approaches promise automated identification of rare or unexpected features in spectral cubes~\cite{sultanov2025anomalyEELS}. These advances open the door to automated tomography, robust intensity quantification, and reliable outlier detection in q-EELS, which in the long run enables faster discovery of new physics.}\\
 
\HCN{Taken together, advances in experimental methods, multi-modal combinations, theoretical modeling and data analysis are positioning q-EELS as a broadly applicable tool for uncovering emergent phenomena in a range of complex materials. }

\section*{Methods (previously unpublished data)}\label{methods}

\subsection*{Sample preparation}
hBN and TMDC samples were exfoliated from bulk crystals (purchased from HQ Graphene) using scotch-tape on silicon wafers with a $285\,\mathrm{nm}$ thick thermal oxide (SiO$_2$/Si) substrate. The dry transfer was done inside a glove box in argon atmosphere, where the optical microscope, the transfer stage and the micromanipulator were controlled from outside. Polydimethylsiloxane (PDMS) and poly(propylene) carbonate (PPC) films were used for the dry transfer due to their favourable viscoelastic and thermoplastic properties. This ensured that high-quality single-layer to few-layer 2D materials were successfully transferred onto holey silicon nitride grids (holey Si$_3$N$_4$ support film of $200\,\mathrm{nm}$ thickness, $1000\,\mathrm{nm}$ pore sizes). This allowed the further analysis, imaging and patterning of the suspended sample to be done in vacuum without background from the support. 

\subsection*{Helium ion microscopy nanopatterning}
A Zeiss Orion Nanofab microscope was used to modify the geometry of the 2D-structures at the nanoscale using He ion beam milling. A nanopatterning and visualization engine (NPVE) software was employed to generate the patterns. In a next step, Stage-o-mat (a special module of FIB-o-mat~\cite{deinhart2021patterning}) was used to place the patterns in such a way that unintended ion irradiation in the suspended sample regions was avoided. Referenced optical control images were used for this purpose. 
An acceleration voltage of 30\,keV, a 11\,mm working distance, a 20\,\textmu m objective aperture, spot size 7, and a beam current of 1.4\,pA were used. The line patterns were created using parallel single line cuts with varying distance rastered using a pitch of 0.25\,nm and 1\,\textmu s dwell time resulting in 20\,pC$/$\textmu m exposure dose. The optimization of the patterning routines was carried out using FIB-o-Mat, which provides complete control over the beam path.

\subsection*{Electron microscopy}
All EEL spectra and images were acquired on a Nion HERMES aberration-corrected high energy resolution monochromated scanning transmission electron microscope at Humboldt-Universität zu Berlin which is equipped with a Dectris ELA  hybrid-pixel direct electron detection camera for recording spectra and diffraction patterns. The microscope was operated at $60\,\mathrm{keV}$ accelerating voltage.

\subsubsection*{Conventional electron energy-loss spectroscopy}
Conventional electron energy-loss maps shown in Fig. were acquired with a convergence angle of 10 mrad and a 2 meV/pixel dispersion in the energy-loss direction. The beam current was $9-10$~pA after monochromation. In order to improve the signal-to-noise, some energy-resolution had to be sacrificed by limiting the monochromation to a factor of 10 to obtain an energy-resolution of $21-35$ meV on the sample.  After acquisition of the spatial EELS maps, the zero-loss-peak (ZLP) was aligned within the Nion Swift software to correct for potential energy shifts. No smoothing was applied to the EEL spectra.

\subsubsection*{Momentum-resolved electron energy-loss spectroscopy in the electron microscope }
All momentum-resolved q-EELS data were acquired at a convergence semi-angle of approximately $1.5$ mrad. A slit aperture was employed to select the high symmetry directions in $k$-space. 
The EELS acquisition with the direct electron detector Dectris ELA allowed to obtain data with high-signal-to-background ratio even at large  \textit{q}.  The energy-momentum maps were aligned for shifts between frames using the Nion Swift software. Next, the ZLP was aligned and centered across the maps. Individual maps of 1s acquisition were then summed to provide the final maps with a total acquisition time of 5-10min. No smoothing of the spectra was employed and all spectra are presented in their raw form. All  $\omega q$ maps are shown in log scale with the momentum information being displayed using fractional units of $|q|$, with $|\Gamma-\Gamma'|=1$ along $\Gamma \rightarrow M$ and $|\Gamma-\Gamma''|=1$ along $\Gamma \rightarrow K$.

%
\section*{Availability of data and materials}
All data that support the findings of this study are available from the corresponding authors upon request.

\bibliography{biblio}

\backmatter

%

%
\bmhead{Acknowledgments}
H.C.N., K.E., and K.H. acknowledge funding by the German Research Foundation, Project No.449639588 (NE 2491/2-1 and HO 5461/5-1) in the framework of the DFG Priority Programme 2244: 2D Materials – Physics of van der Waals [hetero]structures. K.E. and K.H. also acknowledge funding by the German Research Foundation, Project No. 182087777 CRC 951. C.E., F.S.H., Ø.P., and C.T.K acknowledge funding by the Research Council of Norway, Project No. 315330. C.E., F.S.H. and Ø.P. acknowledge the Norwegian Centre for Transmission Electron Microscopy (NORTEM), Project No. 197405, funded by the Research Council of Norway. C.T.K. acknowledges support by the DFG under Project No. 451037016. 
H.C.N. and C.T.K. are grateful for the DFG contribution to the microscopy instrumentation under grant numbers INST 276/721-1 FUGB and INST 276/829-1. We thank Kirill Bolotin and Bianca Höfer (Freie Universität Berlin) for access to the transfer setup and for technical support, respectively. The He ion beam patterning was performed in the Corelab Correlative Microscopy and Spectroscopy at Helmholtz–Zentrum Berlin. The authors want to furthermore acknowledge support by the EU COST action CA 19140 ‘FIT4NANO’ (www.fit4nano.eu).

\bmhead{Author contributions}

\section*{Ethics declarations}

\begin{itemize}

\item Conflict of interest/Competing interests

The authors declare no competing interests.

\item Ethics approval 

Not applicable

\item Consent to participate

Not applicable

\item Consent for publication

Yes

\item Code availability

Not applicable

\end{itemize}

\end{document}